\documentclass[a4paper,11pt]{article}

\usepackage{jheppub} 
\usepackage{amsmath,amssymb,array,calc,rotating,epsfig,psfrag, amscd, datetime}
\usepackage{verbatim}

\newcommand{\nc}{\newcommand}

\definecolor{cardinal}{rgb}{0.6,0,0}
\definecolor{darkgreen}{rgb}{0,0.5,0}
\definecolor{golden}{rgb}{0.92, 0.7, 0}
\definecolor{midnight}{rgb}{0, 0, 0.5}
\definecolor{darkblue}{rgb}{0.2, 0, 0.8}

\nc{\ra}{\rightarrow} 
\nc{\lra}{\leftrightarrow} 
\nc{\Ra}{\Rightarrow} 
\nc{\LRa}{\Leftightarrow} 
\nc{\blp}{{\big (}}
\nc{\brp}{{\big )}}
\nc{\Blp}{{\Big (}}
\nc{\Brp}{{\Big )}}
\nc{\bglp}{{\bigg (}}
\nc{\bgrp}{{\bigg )}}
\nc{\Bglp}{{\Bigg (}}
\nc{\Bgrp}{{\Bigg )}}
\nc{\slb}{{\rm [}}
\nc{\srb}{{\rm ]}}
\nc{\bslb}{{\rm \big [}}
\nc{\bsrb}{{\rm \big ]}}
\nc{\Bslb}{{\rm \Big [}}
\nc{\Bsrb}{{\rm \Big ]}}

\def\al{\alpha}

\def\eps{\epsilon}
\nc{\veps}{\varepsilon}
\def\gam{\gamma}

\def\lam{\lambda}
\def\om{\omega}

\nc{\vphi}{\varphi}
\def\tha{\theta}

\def\sig{\sigma}

\def\Gam{\Gamma}

\def\Lam{\Lambda}
\def\Om{\Omega}
\def\Sig{\Sigma}

\def\coeff#1#2{\relax{\textstyle {#1 \over #2}}\displaystyle}

\nc{\myvspace}{\rule[-1em]{0pt}{2.5em}}
\nc{\bea}{\begin{eqnarray}}
\nc{\eea}{\end{eqnarray}}
\nc{\be}{\begin{equation}}
\nc{\ee}{\end{equation}}
\nc{\barr}{\begin{array}}
\nc{\earr}{\end{array}}
\nc{\cA}{{\cal A}}
\nc{\cB}{ \cal B}

\def\cD{{\cal D}}

\nc{\cF}{{\cal F}}
\nc{\cG}{{\cal G}}

\def\cK{{\cal K}}
\nc{\cL}{{\cal L}}
\nc{\cM}{{\cal M}}
\def\N{{\cal N}}
\def\cN{{\cal N}}

\def\cO{{\cal O}}
\def\cP{{\cal P}}
\nc{\cQ}{{\cal Q}}
\nc{\cR}{{\cal R}}
\def\cS{{\cal S}}

\def\cV{{\cal V}}
\def\cV{{\cal V}}

\def\cZ{{\cal Z}}
\nc{\cQd}{\cQ^{\dagger}}
\nc{\cRd}{\cR^{\dagger}}
\nc{\BB}{{\mathbb B}}
\nc{\CC}{{\mathbb C}}
\nc{\DD}{{\mathbb D}}
\nc{\EE}{{\mathbb E}}
\nc{\FF}{{\mathbb F}}
\nc{\GG}{{\mathbb G}}
\nc{\HH}{{\mathbb H}}
\nc{\JJ}{{\mathbb J}}
\nc{\MM}{{\mathbb M}}
\nc{\RR}{{\mathbb R}}
\nc{\PP}{{\mathbb P}}
\nc{\QQ}{{\mathbb Q}}
\nc{\UU}{{\mathbb U}}
\nc{\ZZ}{{\mathbb Z}}
\nc{\calone}{{\mathbb 1}}
\nc{\half}{\coeff{1}{2}}
\nc{\quarter}{\coeff{1}{4}}
\nc{\del}{\partial}

\nc{\delbar}{\bar\partial}
\nc{\thalf}{\frac{t}{2}}
\nc{\Spin}{\operatorname{Spin}}
\nc{\SO}{\operatorname{SO}}

\nc{\Sp}{{\rm Sp}}
\nc{\com}[2]{{ \left[ #1, #2 \right] }}
\nc{\acom}[2]{{ \left\{ #1, #2 \right\} }}
\nc{\rr}{\rightarrow}
\nc{\p}{\partial}
\nc{\LT}{{\LL_\T}}
\nc{\Tr}{{\rm Tr}}
\nc{\tr}{{\rm tr}}
\nc{\Adag}{A^{\dagger}}
\nc{\AdagI}{A^{\dagger I}}
\nc{\AdagJ}{A^{\dagger J}}
\nc{\AdagK}{A^{\dagger K}}
\nc{\AdagL}{A^{\dagger L}}
\nc{\AdagM}{A^{\dagger M}}
\nc{\Bdag}{B^{\dagger}}
\nc{\BdagI}{B^{\dagger}_I}
\nc{\BdagJ}{B^{\dagger}_J}
\nc{\BdagK}{B^{\dagger}_K}
\nc{\BdagL}{B^{\dagger}_L}
\nc{\BdagM}{B^{\dagger}_M}
\nc{\Cdag}{C^{\dagger}}
\nc{\CdagI}{C^{\dagger I}}
\nc{\CdagJ}{C^{\dagger J}}
\nc{\CdagK}{C^{\dagger K}}
\nc{\Ddag}{D^{\dagger}}
\nc{\DdagI}{D^{\dagger I}}
\nc{\DdagJ}{D^{\dagger J}}
\nc{\DdagK}{D^{\dagger K}}
\nc{\ttha}{\tilde{\theta}}
\nc{\ttau}{\tilde{\tau}}
\nc{\tTha}{\tilde{\Theta}}
\nc{\tphi}{\tilde{\phi}}
\nc{\tsig}{\tilde{\sig}}
\nc{\tom}{\widetilde{\om}}
\nc{\tOm}{\widetilde{\Om}}
\nc{\tlam}{\widetilde{\lam}}
\nc{\tLam}{\tilde{\Lam}}
\nc{\tSig}{\widetilde{\Sig}}
\nc{\tPhi}{\tilde{\Phi}}
\nc{\tPhibar}{\ol{\tPhi}}
\nc{\tPi}{\widetilde{\Pi}}
\nc{\tpsi}{\widetilde{\psi}}
\nc{\tPsi}{\tilde{\Psi}}
\nc{\tgam}{\widetilde{\gam}}
\nc{\tGam}{\widetilde{\Gam}}
\nc{\tzeta}{\tilde{\zeta}}
\nc{\tZeta}{\tilde{\Zeta}}
\nc{\teta}{\widetilde{\eta}}
\nc{\teps}{\tilde{\eps}}
\nc{\tveps}{\tilde{\veps}}
\nc{\tEta}{\tilde{\Eta}}
\nc{\tchi}{\tilde{\chi}}
\nc{\tChi}{\tilde{\Chi}}
\nc{\txi}{\tilde{\xi}}
\nc{\tXi}{\widetilde{\Xi}}
\nc{\tnu}{\tilde{\nu}}
\nc{\tmu}{\tilde{\mu}}

\nc{\tb}{\tilde b}
\nc{\tc}{\tilde c}
\nc{\te}{\tilde e}
\nc{\tf}{\tilde f}
\nc{\tg}{\tilde g}
\nc{\ti}{\tilde i}
\nc{\tj}{\tilde j}
\nc{\tk}{\tilde k}
\nc{\tl}{\tilde l}
\nc{\tm}{\tilde m}
\nc{\tn}{\tilde n}
\nc{\tp}{\tilde{p}}
\nc{\tq}{\widetilde{q}}
\nc{\ts}{{\tilde s}}
\nc{\tu}{{\tilde u}}
\nc{\tv}{{\tilde v}}
\nc{\tw}{{\tilde w}}
\nc{\tx}{{\tilde x}}
\nc{\ty}{{\tilde y}}
\nc{\tz}{\tilde z}
\nc{\tA}{{\widetilde A}}
\nc{\tAbar}{{\ol \tA}}
\nc{\tB}{{\widetilde B}}
\nc{\tC}{{\widetilde C}}
\nc{\tD}{{\widetilde D}}
\nc{\tE}{{\widetilde E}}
\nc{\tF}{{\widetilde F}}
\nc{\tG}{{\widetilde G}}
\nc{\tH}{{\widetilde H}}
\nc{\tJ}{{\widetilde J}}
\nc{\tJbar}{{\ol {\tilde J}}}
\nc{\tK}{{\widetilde K}}
\nc{\tL}{{\widetilde L}}
\nc{\tcL}{{\widetilde \cL}}
\nc{\tM}{{\widetilde M}}
\nc{\tN}{{\widetilde N}}
\nc{\tcN}{{\widetilde \cN}}
\nc{\tP}{{\widetilde P}}
\nc{\tQ}{{\widetilde Q}}
\nc{\tR}{{\widetilde R}}
\nc{\tS}{\widetilde{S}}
\nc{\tT}{\widetilde{T}}
\nc{\tU}{\widetilde{U}}
\nc{\tV}{\widetilde{V}}
\nc{\tW}{\widetilde{W}}
\nc{\tcF}{\widetilde{{\cal F}}}
\nc{\tX}{\widetilde{X}}
\nc{\tY}{\widetilde{Y}}
\nc{\tcZ}{\tilde{\cZ}}
\nc{\tcZbar}{\ol{\tcZ}}

\nc{\ha}{\hat a}
\nc{\hb}{\hat b}
\nc{\hc}{\widehat c}
\nc{\hd}{\widehat d}
\nc{\he}{\widehat e}
\nc{\hf}{\widehat f}
\nc{\hg}{\widehat g}
\nc{\hh}{\widehat h}
\nc{\hm}{\widehat m}
\nc{\hn}{\widehat n}
\nc{\hp}{\widehat p}
\nc{\hr}{\widehat r}
\nc{\hs}{\widehat s}
\nc{\hv}{\widehat v}
\nc{\hw}{\widehat w}
\nc{\hx}{\widehat x}
\nc{\hy}{\widehat y}
\nc{\hz}{\widehat z}
\nc{\zhat}{\hat z}
\nc{\hA}{\widehat{A}}
\nc{\hB}{\widehat{B}}
\nc{\hC}{\widehat{C}}
\nc{\hD}{\widehat{D}}
\nc{\hE}{\widehat{E}}
\nc{\hF}{\widehat{F}}
\nc{\hcF}{\widehat{\cF}}
\nc{\hG}{\widehat{G}}
\nc{\hH}{\widehat{H}}
\nc{\hJ}{\widehat{J}}
\nc{\hK}{\widehat{K}}
\nc{\hL}{\widehat{L}}
\nc{\hcL}{\widehat{\cL}}
\nc{\hM}{\widehat M}
\nc{\hcM}{\widehat{\cM}}
\nc{\hN}{\widehat{N}}
\nc{\hO}{\widehat{O}}
\nc{\hP}{\widehat{P}}
\nc{\hQ}{\widehat{Q}}
\nc{\hcR}{\widehat{\cR}}
\nc{\hR}{\widehat{R}}
\nc{\hS}{\widehat{S}}
\nc{\hcS}{\widehat{\cS}}
\nc{\hT}{\widehat{T}}
\nc{\hU}{\widehat{U}}
\nc{\hV}{\widehat V}
\nc{\hcV}{\widehat \cV}
\nc{\hX}{\widehat X}

\nc{\heta}{\widehat{\eta}}
\nc{\hal}{\widehat \alpha}
\nc{\hphi}{\widehat{\phi}}
\nc{\hkap}{\hat{\kappa}}
\nc{\hchi}{\widehat{\chi}}
\nc{\hpsi}{\widehat{\psi}}
\nc{\hgam}{\widehat{\gam}}
\nc{\hPhi}{\hat{\Phi}}
\nc{\hPsi}{\hat{\Psi}}
\nc{\hGam}{\hat{\Gam}}
\nc{\omhat}{\widehat{\om}}
\nc{\htha}{\hat{\tha}}
\nc{\hrho}{\widehat{\rho}}
\nc{\hdel}{\widehat{\del}}

\nc{\w}{\wedge}

\nc{\vb}{\vec b}
\nc{\vc}{\vec c}
\nc{\vd}{\vec d}
\nc{\ve}{\vec e}
\nc{\vf}{\vec f}
\nc{\vg}{\vec g}
\nc{\vh}{\vec h}
\nc{\vp}{\vec p}
\nc{\vq}{\vec q}
\nc{\vr}{\vec r}
\nc{\vs}{\vec s}
\nc{\vv}{\vec v}
\nc{\vw}{\vec w}
\nc{\vx}{\vec x}
\nc{\vy}{\vec y}
\nc{\vz}{\vec z}

\nc{\vB}{\vec B}
\nc{\vC}{\vec C}
\nc{\vD}{\vec D}
\nc{\vE}{\vec E}
\nc{\vF}{\vec F}
\nc{\vG}{\vec G}
\nc{\vH}{\vec H}
\nc{\vP}{\vec P}
\nc{\vQ}{\vec Q}
\nc{\vR}{\vec R}
\nc{\vS}{\vec S}
\nc{\vV}{\vec V}
\nc{\vW}{\vec W}
\nc{\vX}{\vec X}
\nc{\vY}{\vec Y}
\nc{\vZ}{\vec Z}

\nc{\ol}{\overline}
\nc{\abar}{\ol{a}}
\nc{\bbar}{\ol{b}}
\nc{\cbar}{\ol{c}}
\nc{\dbar}{\ol{d}}
\nc{\ebar}{\ol{e}}
\nc{\fbar}{\ol{f}}
\nc{\ibar}{\ol{\imath}}
\nc{\jbar}{\ol{\jmath}}
\nc{\kbar}{\ol{k}}
\nc{\lbar}{\ol{l}}
\nc{\mbar}{\ol{m}}
\nc{\nbar}{\ol{n}}
\nc{\pbar}{\ol{p}}
\nc{\qbar}{\ol{q}}
\nc{\rbar}{\ol{r}}
\nc{\sbar}{\ol{s}}
\nc{\ubar}{\ol{u}}
\nc{\vbar}{\ol{v}}
\nc{\wbar}{\ol{w}}
\nc{\xbar}{\ol{x}}
\nc{\ybar}{\ol{y}}
\nc{\zbar}{\ol{z}}

\nc{\Abar}{\ol{A}}
\nc{\Bbar}{\ol{B}}
\nc{\Cbar}{\ol{C}}
\nc{\Dbar}{\ol{D}}
\nc{\Ebar}{\ol{E}}
\nc{\Fbar}{\ol{F}}
\nc{\Jbar}{\ol{J}}
\nc{\Kbar}{\ol{K}}
\nc{\Lbar}{\ol{L}}
\nc{\cLbar}{\ol{\cL}}
\nc{\Mbar}{\ol{M}}
\nc{\Nbar}{\ol{N}}
\nc{\Pbar}{\ol{P}}
\nc{\Qbar}{\ol{Q}}
\nc{\Rbar}{\ol{R}}
\nc{\Sbar}{\ol{S}}
\nc{\Tbar}{\ol{T}}
\nc{\Ubar}{\ol{U}}
\nc{\Vbar}{\ol{V}}
\nc{\cVbar}{\ol{\cV}}
\nc{\Wbar}{\ol{W}}
\nc{\Xbar}{{\overline X}}
\nc{\Ybar}{{\overline Y}}
\nc{\Zbar}{{\overline Z}}
\nc{\cZbar}{{\overline \cZ}}

\nc{\epsbar}{\ol{\epsilon}}
\nc{\lambar}{\ol{\lambda}}
\nc{\kapbar}{\ol{\kappa}}
\nc{\zetabar}{\ol{\zeta}}
\nc{\Zetabar}{\ol{\Zeta}}
\nc{\taubar}{\ol{\tau}}
\nc{\Taubar}{\ol{\Tau}}
\nc{\psibar}{\ol{\psi}}
\nc{\Psibar}{\ol{\Psi}}
\nc{\tpsibar}{\ol{\tpsi}}
\nc{\tPsibar}{\ol{\tPsi}}
\nc{\phibar}{\ol{\phi}}
\nc{\Phibar}{\ol{\Phi}}
\nc{\chibar}{\ol{\chi}}
\nc{\mubar}{\ol{\mu}}
\nc{\nubar}{\ol{\nu}}
\nc{\rhobar}{\ol{\rho}}
\nc{\ombar}{\ol{\om}}
\nc{\Ombar}{\ol{\Om}}
\nc{\Deltabar}{\ol{\Delta}}
\nc{\Thetabar}{\ol{\Theta}}
\nc{\xibar}{\ol{\xi}}
\nc{\Xibar}{\ol{\Xi}}

\nc{\Dthbar}{\ol{\rm D3}}

\nc{\gdot}{\dot{g}}
\nc{\pdot}{\dot{p}}
\nc{\qdot}{\dot{q}}
\nc{\rdot}{\dot{r}}
\nc{\sdot}{\dot{s}}
\nc{\tdot}{\dot{t}}
\nc{\udot}{\dot{u}}
\nc{\vdot}{\dot{v}}
\nc{\wdot}{\dot{w}}
\nc{\xdot}{\dot{x}}
\nc{\xddot}{\ddot{x}}
\nc{\ydot}{\dot{y}}
\nc{\zdot}{\dot{z}}
\nc{\yddot}{\ddot{y}}

\nc{\Udot}{\dot{U}}
\nc{\Vdot}{\dot{V}}
\nc{\Wdot}{\dot{W}}

\nc{\taudot}{\dot{\tau}}
\nc{\phidot}{\dot{\phi}}
\nc{\psidot}{\dot{\psi}}
\nc{\sinp}{s_{\phi}}
\nc{\cosp}{c_{\phi}}
\nc{\tanp}{t_{\phi}}
\nc{\spone}{s_{\phi_1}}
\nc{\cpone}{c_{\phi_1}}
\nc{\tpone}{t_{\phi_1}}
\nc{\sptwo}{s_{\phi_2}}
\nc{\cptwo}{c_{\phi_2}}
\nc{\tptwo}{t_{\phi_2}}
\nc{\spth}{s_{\phi_3}}
\nc{\cpth}{c_{\phi_3}}
\nc{\tpth}{t_{\phi_3}}
\nc{\calp}{c_{\al}}
\nc{\salp}{s_{\al}}

\nc{\csch}{{\rm csch}}
\nc{\sech}{{\rm sech}}

\nc{\cothzlami}{\coth(z-\lam_i)}
\nc{\coshzlami}{\cosh(z-\lam_i)}
\nc{\sinhzlami}{\sinh(z-\lam_i)}

\nc{\cothzlamj}{\coth(z-\lam_j)}
\nc{\coshzlamj}{\cosh(z-\lam_j)}
\nc{\sinhzlamj}{\sinh(z-\lam_j)}

\nc{\cothlamij}{\coth(\lam_i-\lam_j)}
\nc{\coshlamij}{\cosh(\lam_i-\lam_j)}
\nc{\sinhlamij}{\sinh(\lam_i-\lam_j)}

\nc{\bah}{{\mathbf {\hat{A}}}}
\nc{\bX}{{\mathbf X}}
\nc{\ba}{{\bf a}}
\nc{\bb}{{\bf b}}
\nc{\bc}{{\bf c}}
\nc{\bd}{{\bf d}}
\nc{\bg}{{\bf g}}
\nc{\bk}{{\bf k}}
\nc{\bl}{{\bf l}}
\nc{\bm}{{\bf m}}
\nc{\bn}{{\bf n}}
\nc{\bo}{{\bf o}}
\nc{\bp}{{\bf p}}
\nc{\bq}{{\bf q}}
\nc{\br}{{\bf r}}
\nc{\bs}{{\bf s}}
\nc{\bt}{{\bf t}}
\nc{\bu}{{\bf u}}
\nc{\bv}{{\bf v}}
\nc{\bw}{{\bf w}}
\nc{\bx}{{\bf x}}
\nc{\by}{{\bf y}}
\nc{\bz}{{\bf z}}
\nc{\bom}{{\bf \om}}
\nc{\bombar}{{\mathbf \ombar}}
\nc{\bPhi}{{\bf \Phi}}

\nc{\rma}{{\rm a}}
\nc{\rmb}{{\rm b}}
\nc{\rmc}{{\rm c}}
\nc{\rmd}{{\rm d}}
\nc{\rmg}{{\rm g}}
\nc{\rk}{{\rm k}}
\nc{\rml}{{\rm l}}
\nc{\rmm}{{\rm m}}
\nc{\rmn}{{\rm n}}
\nc{\rmo}{{\rm o}}
\nc{\rmp}{{\rm p}}
\nc{\rmq}{{\rm q}}
\nc{\rmr}{{\rm r}}
\nc{\rms}{{\rm s}}
\nc{\rmt}{{\rm t}}
\nc{\rmu}{{\rm u}}
\nc{\rmv}{{\rm v}}
\nc{\rmw}{{\rm w}}
\nc{\rmx}{{\rm x}}
\nc{\rmy}{{\rm y}}
\nc{\rmz}{{\rm z}}

\nc{\dal}{\dot{\al}}
\nc{\thadot}{\dot{\tha}}
\nc{\thab}{\bar{\theta}}
\nc{\thal}{\theta^{\al}}
\nc{\thdal}{\bar{\theta}^{\dal}}

\nc{\thsigthm}{\tha \sigma^m \thab}
\nc{\thsigthn}{\tha \sigma^n \thab}

\nc{\Dal}{D_{\al}}
\nc{\Ddal}{\bar{D}_{\dal}}
\nc{\CDal}{{\cal D}_{\al}}
\nc{\CDdal}{\bar{\cal D}_{\dal}}

\nc{\eq}[1]{(\ref{#1})}
\nc{\non}{\nonumber}
\nc{\Fzero}{F_{(0)}}
\nc{\Ftwo}{F_{(2)}}
\nc{\Ffour}{F_{(4)}}
\nc{\Fone}{F_{(1)}}
\nc{\Fthree}{F_{(3)}}
\nc{\Ffive}{F_{(5)}}
\nc{\Fn}{F_{(n)}}
\nc{\Fp}{F_{(p)}}

\nc{\tFzero}{\tF_{(0)}}
\nc{\tFtwo}{\tF_{(2)}}
\nc{\tFfour}{\tF_{(4)}}
\nc{\tFone}{\tF_{(1)}}
\nc{\tFthree}{\tF_{(3)}}
\nc{\tFfive}{\tF_{(5)}}
\nc{\tFn}{\tF_{(n)}}
\nc{\tFp}{\tF_{(p)}}

\nc{\Czero}{C_{(0)}}
\nc{\Ctwo}{C_{(2)}}
\nc{\Cfour}{C_{(4)}}
\nc{\Cone}{C_{(1)}}
\nc{\Cthree}{C_{(3)}}
\nc{\Cfive}{C_{(5)}}
\nc{\Cn}{C_{(n)}}

\nc{\equ}{{\rm eq}}
\def\Im{{\rm Im \hspace{0.5mm} }}

\def\Re{{\rm Re \hspace{0.5mm}}}

\nc{\vol}{{\rm vol}}
\nc{\Ainf}{A_{\infty}}
\nc{\End}{{\rm End}}
\nc{\Ext}{{\rm Ext}}
\nc{\IIB}{{\rm IIB}}
\nc{\Ad}{{\rm Ad}}
\nc{\IIA}{{\rm IIA}}
\nc{\AdS}{{\rm AdS}}
\nc{\CFT}{{\rm CFT}}
\nc{\diag}{{\rm diag}}
\nc{\Log}{{\rm Log}}
\nc{\Dslash}{\ensuremath \raisebox{0.025cm}{\slash}\hspace{-0.32cm} D}
\nc{\cDslash}{\ensuremath \raisebox{0.025cm}{\slash}\hspace{-0.32cm} \cD}
\nc{\omslash}{\om\!\!\!/}
\nc{\no}{\!:\!\!}
\nc{\ointdz}{\oint\frac{dz}{2\pi i}}
\nc{\ointdzone}{\oint\frac{dz_1}{2\pi i}}
\nc{\ointdztwo}{\oint\frac{dz_2}{2\pi i}}
\nc{\ointdzb}{\oint\frac{d\zbar}{2\pi i}}
\nc{\ointdzbone}{\oint\frac{d\zbar_1}{2\pi i}}
\nc{\ointdzbtwo}{\oint\frac{d\zbar_2}{2\pi i}}
\nc{\dz}{\frac{dz}{2\pi i}}
\nc{\dzb}{\frac{d\zbar}{2\pi i}}
\nc{\bpm}{\begin{pmatrix}}
\nc{\epm}{\end{pmatrix}}
 \nc{\bitem}{\begin{itemize}}
 \nc{\eitem}{\end{itemize}}
 \nc{\exercise}{\vskip 2mm \noindent {\bf Exercise:}}
 \nc{\definition}{\vskip 2mm \noindent {\bf Definition:}}
 
\newcommand\dd{\mathrm{d}}

\newcommand{\nn}{\nonumber \\ {} }
\newcommand\ex{\mathrm{e}}
\newcommand\ii{\mathrm{i}}
\newcommand\D{\mathrm{D}}
\newcommand\qqq{\qquad\qquad}
\title{BPS M5-branes as Defects for the 3d-3d Correspondence}

\author[a]{Ibrahima Bah}
\author[b,c]{Maxime Gabella}
\author[d,e]{Nick Halmagyi}

\affiliation[a]{Department of Physics and Astronomy, University of Southern California, Los Angeles, CA 90089, USA}
\affiliation[b]{Department of Mathematics, University of Hamburg, Bundesstra\ss e 55, 20146 Hamburg, Germany}
\affiliation[c]{DESY, Theory Group, Notkestra\ss e 85, Building 2a, 22607 Hamburg, Germany}
\affiliation[d]{Sorbonne Universit\'es, UPMC Paris 06, UMR 7589, LPTHE, 75005, Paris, France}
\affiliation[e]{CNRS, UMR 7589, LPTHE, 75005, Paris, France}

\emailAdd{bah@usc.edu}
\emailAdd{maxime.gabella@uni-hamburg.de}
\emailAdd{halmagyi@lpthe.jussieu.fr}

\abstract{We study supersymmetric probe M5-branes in the AdS$_4$ solution that arises from M5-branes wrapped on a hyperbolic 3-manifold $M_3$. 
This amounts to introducing internal defects within the framework of the 3d-3d correspondence.
The BPS condition for a probe M5-brane extending along all of AdS$_4$ requires it to wrap a surface embedded in an $S^2$-fibration over~$M_3$. We find that the projection of this surface to $M_3$ can be either a geodesic or a tubular surface around a geodesic. 
These configurations preserve an extra $U(1)$ symmetry, in addition to the one corresponding to the R-symmetry of the dual 3d $\cN=2$ gauge theory.
BPS M2-branes can stretch between M5-branes wrapping geodesics. We interpret the addition of probe M5-branes on a closed geodesic in terms of conformal Dehn surgery.}

\begin{document} 

\maketitle

%%%%%%%%%%%%%%%%%%%%%%%%%%%%%%%%%
%%%%%%%%%%%%%%%%%%%%%%%%%%%%%%%%%
%%%%%%%%%%%%%%%%%%%%%%%%%%%%%%%%%
\section{Introduction}
%%%%%%%%%%%%%%%%%%%%%%%%%%%%%%%%%

The 3d-3d correspondence associates a 3d $\cN=2$ supersymmetric gauge theory to a 3-manifold $M_3$~\cite{Dimofte:2011ju,Cecotti:2011iy,Dimofte:2013iv,Lee:2013ida,Cordova:2013cea}.
This theory can be thought of as the partially twisted compactification of the 6d $(2,0)$  superconformal field theory. 
It describes the low-energy effective theory on the worldvolume of a stack of $N$ M5-branes wrapping $\RR^{1,2}\times M_3$ in the 11d spacetime $\RR^{1,4}\times \text{CY}_3$, where $M_3$ is embedded in the Calabi-Yau 3-fold CY$_3$ as a special Lagrangian submanifold.

In this paper, we are interested in the large $N$ limit of this setup with CY$_3$ taken to be the cotangent bundle $T^*M_3$.
The stack of $N$ M5-branes backreacts on the geometry and gives rise in the IR to an AdS$_4$ solution of 11d supergravity.
We sketch this process as follows:
\bea
\RR^2 \times \underbrace{\RR^{1,2} \times \big(M_3}_{N~\text{M5-branes}} \leftarrow \RR^3  \big) \qquad \stackrel{N \gg 1} {\Longrightarrow} \qquad 
S^1_\text{R} \times \text{AdS}_4 \times \big(M_3 \leftarrow S^2\big) \times I_\rho~.
\eea
The radial coordinates of the factor $\RR^2$ and of the fibers $\RR^3$ in $T^*M_3 = ( M_3 \leftarrow \RR^3)$ combine in the large $N$ limit to give the radial coordinate of AdS$_4$ and a coordinate $\rho$ on the interval $I_\rho$.
Accordingly, the cotangent bundle $T^*M_3$ is replaced by the unit cotangent bundle $T_1^*M_3 = (M_3 \leftarrow S^2)$.
The circle $S^1_\text{R}$ corresponds to the $U(1)$ R-symmetry of the dual 3d $\cN=2$ SCFT.
This circle shrinks to a point at the origin of the interval $I_\rho$, while the sphere $S^2$ shrinks at the other end, so that together they have the topology of an $S^4$.
This solution is in fact the M-theory uplift of a compactification of 7d gauged supergravity on AdS$_4\times \HH^3$ discovered by Pernici and Sezgin~\cite{Pernici:1984nw} (reviewed in section~\ref{sec:AdS4}).
Hyperbolic 3-space $\HH^3$ can be made compact by quotienting by a discrete subgroup $\Gamma\subset PSL(2,\CC)$:
\bea
M_3=\HH^3/\Gamma ~.
\eea

In section~\ref{secM5}, we consider adding supersymmetric probe M5-branes to the Pernici-Sezgin solution, in such a way that the superconformal symmetry of the dual theory is preserved. 
This corresponds to studying the space of 3d $\cN=2$ SCFTs arising from M5-branes on 3-manifolds
(some of our inspiration came from the interpretation of probe M5-branes as punctures on a Riemann surface associated with a 4d $\cN=2$ SCFT~\cite{Gaiotto:2009gz} or with a 4d $\cN=1$ SCFT~\cite{Bah:2013wda}).
We thus require that the probe M5-branes extend along all of AdS$_4$ and do not break the $U(1)$ R-symmetry corresponding to $S^1_\text{R}$. 
The BPS condition coming from $\kappa$-symmetry then implies that a supersymmetric M5-brane wraps a surface embedded in the unit cotangent bundle $T_1^*M_3$.
The first type of BPS embedding that we find is an M5-brane wrapping a geodesic in $M_3$ as well as a great circle on the $S^2$-fiber. 
This was anticipated, since line defects in 3-manifolds (such as the knot of a knot complement) are the analogues of punctures on a Riemann surface.
Perhaps more unexpectedly, we also find that a BPS M5-brane can warp a tubular surface in $M_3$, for instance a torus or an annulus.

These BPS configurations should be imagined as descending from a probe M5-brane wrapping a special Lagrangian submanifold of $T^*M_3$ in the UV. It appears that this submanifold is the conormal bundle $N^*L$ of a submanifold $L\subset M_3$, which can be a geodesic or a tube. Note that $N^*L$ is automatically Lagrangian in $T^*M_3$. 
Intersecting branes on conormal bundles have been used in various contexts to build knot complements (see for example~\cite{Ooguri:1999bv,Witten:2011zz,Aganagic:2013jpa}).
In the IR, the probe M5-brane wraps the unit conormal bundle $N_1^*L \subset T^*_1M_3$.
The BPS embeddings that we found for a supersymmetric probe M5-brane are thus of the form
\bea
\text{\bf BPS M5-brane}:\quad \text{AdS}_4 \times N^*_1 L\times \{\rho=0\} \quad \subset \quad 
 \text{AdS}_4 \times T^*_1 M_3 \times I_\rho \times S^1_\text{R}~.
\eea
Interestingly, the unit conormal bundles $N_1^*L$ that we obtain are flat (with the topology of a cylinder), which means that these configurations preserve an extra $U(1)$ symmetry, in addition to the $U(1)$ R-symmetry corresponding to $S^1_\text{R}$.

In section~\ref{secM2}, we study BPS embeddings for supersymmetric probe M2-branes that correspond to BPS operators in the dual SCFT.
We focus in particular on probe M2-branes ending on probe M5-branes along two spacetime dimensions. We find BPS M2-branes stretching between M5-branes on geodesics contained in the same geodesic surface, and M2-branes stretching between great circles wrapped by M5-branes on an $S^2$-fiber. In contrast, we did not find M2-branes ending on M5-branes wrapping surfaces in $M_3$.
We take this as an indication that the two types of BPS M5-branes will play rather different roles in the dual 3d theories.

In the BPS calculations that we perform, we neglect the quotient by $\Gamma$ that produces a closed 3-manifold $M_3 = \HH^3/\Gamma$, and study embeddings in $T^*_1\HH^3$.
In section~\ref{hyperbolicinterpretation}, we present a potential interpretation of our results in the actual manifold $M_3$.
We relate the addition of probe M5-branes on a closed geodesic to conformal Dehn surgery, which consists of excising a solid torus from $M_3$, twisting it, and gluing it back in.
The number of coincident M5-branes should correspond to the amount of twisting, and taking it to be very large would produce a non-compact hyperbolic 3-manifold with a cusp, for example a knot complement.

%%%%%%%%%%%%%%%%%%%%%%%%%%%%%%%%%
%%%%%%%%%%%%%%%%%%%%%%%%%%%%%%%%%
\section{Pernici-Sezgin AdS$_4$ solution} \label{sec:AdS4}
%%%%%%%%%%%%%%%%%%%%%%%%%%%%%%%%%

We start by reviewing the AdS$_4$ solution of M-theory that we will be considering in this paper.
This solution is the 11d uplift of a 7d gauged supergravity solution originally found by Pernici and Sezgin (PS)~\cite{Pernici:1984nw}, and was subsequently rediscovered in~\cite{Acharya:2000mu,Gauntlett:2000ng} from a study of wrapped M5-branes.
It is of the form AdS$_4 \times Y_7$, 
where $Y_7$ is an $S^4$-fibration over a hyperbolic 3-manifold $M_3$.

This solution was shown in~\cite{Gauntlett:2006ux} to arise as a special case of a general class of $\N=2$ supersymmetric AdS$_4$ geometries describing the near-horizon limit of M5-branes wrapping a special Lagrangian 3-cycle in a Calabi-Yau 3-fold
(the generality of this class was proven in~\cite{Gabella:2012rc}).
The metric for this class of solutions takes the form
\bea \label{11dmetricAdS4}
\dd s^2_{11} &=& \lambda^{-1} \dd s^2 (\text{AdS}_4) + \dd s_4^2 (\cM_{SU(2)}) + \hat w \otimes \hat w + \frac{\lambda^2}{16} \left(\frac{\dd \rho^2}{1-\lambda^3\rho^2} + \rho^2 \dd\psi^2 \right)~,
\eea
where $\lambda$ is the warp factor, $\cM_{SU(2)}$ is a 4d space with $SU(2)$-structure, $\hat w$ is a one-form, $\rho$ is an interval coordinate, and $\psi\in [0,2\pi]$ is a coordinate on a circle $S^1_\text{R}$.
The Killing vector field $\del/\del \psi$ is dual to the $U(1)$ R-symmetry.
The supersymmetry conditions reduce to a system of differential equations involving the standard two-forms $\{J_1,J_2,J_3\}$ defining the $SU(2)$-structure.
The four-form flux is given by 
\bea\label{G4flux}
G_4 &=& \frac1{4} \dd\psi \wedge \dd \left( \lambda^{-1/2} \sqrt{1-\lambda^3\rho^2} J_3\right)~.
\eea

The PS solution was reproduced in section 9.5 of~\cite{Gauntlett:2006ux} by making the following ansatz:
\bea\label{metDYe}
\lambda &=& \lambda(\rho)~, \nn
 \dd s_4^2 ( \cM_{SU(2)}) + \hat w \otimes \hat w &=& f^2(\rho) \D  Y^a \D  Y^a + g^2(\rho) e^a e^a~,
\eea
where $Y^a$, $a=1,2,3$, are constrained coordinates on an $S^2$ satisfying $Y^aY^a=1$, and $e^a$ are vielbeins for a 3-manifold $M_3$. 
The covariant derivative is defined as
\bea
\D Y^a &=& \dd Y^a + \omega^a{}_b Y^b~,
\eea
with $\omega^{ab}$ the spin connection of $M_3$.
The ansatz for the structure is
\bea\label{ansatzstr}
\hat w &=& g Y^a e^a~, \nn
J_1 &=& fg \D Y^a \wedge e^a~, \nn
J_2 &=& fg \epsilon^{abc} Y^a \D Y^b \wedge e^c~, \nn
J_3 &=& \tfrac12 \epsilon^{abc} Y^a \left( f^2 \D Y^b \wedge \D Y^c - g^2 e^b\wedge e^c \right)~.
\eea
The supersymmetry conditions then determines the functions as
\bea
\lambda^3 = \frac{2}{ 8+\rho^2}~, \qqq 
f = \frac{\sqrt{1- \lambda^3 \rho^2}}{2 \lambda^{1/2}}~, \qqq 
g =  \frac{1}{ 2^{3/2} \lambda^{1/2}}~,
\eea
and imply that $M_3$ is a hyperbolic 3-manifold and that the coordinates $Y^a$ together with $\rho$ and $\psi$ build up an $S^4$.
The hyperbolic 3-manifold $M_3$ can be expressed as a quotient of hyperbolic 3-space $\HH^3$ by an discrete subgroup $\Gamma\subset PSL(2,\CC)$, that is $M_3 = \HH^3/\Gamma$ (see appendix~\ref{hyperbolicgeom} for a small review on hyperbolic 3-manifolds). 
The metric on $M_3$ is normalized such that the Ricci scalar is $R=-3$, so we take 
\bea
\dd s^2 (M_3) =  2 \dd s^2(\HH^3)  =  2\frac{\dd x^2 + \dd y^2 + \dd z^2}{z^2} ~, \qqq z > 0~. 
\eea
The spin connection on $M_3$ has the non-zero components $\omega_{31} = \dd x/z$ and $\omega_{32} = \dd y /z$.
The metric of the PS solution finally reads
\bea \label{PSmet}
\dd s^2_{11} &=&  \frac14 \lambda^{-1}  \bigg[4 \dd s^2 (\text{AdS}_4) +  \frac{\dd x^2 + \dd y^2 + \dd z^2}{z^2}  \nn
&& \qqq +  \frac{8 - \rho^2 }{8+\rho^2 } \D Y^a \D Y^a   +  \frac12\left( \frac{\dd \rho^2}{8-\rho^2}  + \frac{\rho^2}{8+\rho^2} \dd \psi^2\right) \bigg]~.
\eea

The holographic free energy can be easily calculated as
\bea
\cF = \frac{\pi }{2G_4^\text{N}} = \frac{\pi^5 \text{Vol}(M_3)}{2(2\pi l_p)^9}~,
\eea
where $G_4^{\text{N}}$ is the effective 4d Newton constant obtained by dimensional reduction (see~\cite{Gabella:2012rc} for more detail).
The quantization of the four-form flux~\eqref{G4flux},
\bea
N = \frac{1}{(2\pi \ell_p)^3} \int_{X_4} G_4= \frac{1}{8\pi\ell_p^3  }~,
\eea
with $X_4\subset Y_7$ transverse to $M_3$, then gives the expected $N^3$-scaling:
\bea
\cF = \frac{N^3}{3\pi} \text{Vol}(M_3)~.
\eea
Note that the free energy corresponding to SCFTs on a squashed 3-sphere $S^3_b$ with squashing parameter $b$ is simply given by $\cF_b = (b+1/b)^2 \cF/4$~\cite{Martelli:2011fu,Gang:2014qla}.

%%%%%%%%%%%%%%%%%%%%
\section{Supersymmetric probe M5-branes}\label{secM5}
%%%%%%%%%%%%%%%%%%%%

In this section, we study a supersymmetric probe M5-brane that preserves the superconformal symmetries of the dual 3d $\cN=2$ SCFT.
This implies that it should extend along all of AdS$_4$, and that the remaining two internal directions should wrap a 2d submanifold of~$Y_7$ in such a way as to preserve the $U(1)$ R-symmetry. 
There can be no three-form flux on the worldvolume of the M5-brane, since it would have to extend along at least one AdS$_4$ dimension, thus breaking the conformal symmetry. 
We will see that the BPS condition arising from $\kappa$-symmetry imposes that the probe M5-brane is located at the origin of the interval $I_\rho$, where $S^1_\text{R}$ shrinks to a point, and that it is calibrated by the two-form~$J_2$.
The BPS configurations that we will find describe an M5-brane wrapping the unit conormal bundle $N^*_1L \subset T^*_1M_3$ of a submanifold $L\subset M_3$.
This submanifold $L$ can be a geodesic curve, but also a surface that is equidistant from a point at infinity or from a geodesic curve.
These BPS embeddings preserve an extra $U(1)$ symmetry, in addition to the $U(1)$ R-symmetry.

%%%%%%%%%%%%%%%%%%%%
\subsection{BPS condition}\label{secBPSM5}
%%%%%%%%%%%%%%%%%%%%

The requirement of $\kappa$-symmetry leads to a BPS bound on a supersymmetric probe M5-brane \cite{Becker:1995kb} (see also~\cite{Martelli:2003ki}).
A configuration that preserves some supersymmetry satisfies $\cP_-\epsilon=0$, where
$\epsilon$ is a Majorana spinor of 11d supergravity satisfying the Killing spinor equation, and $\cP_-$ is a $\kappa$-symmetry projector.
Explicitly, we have $\cP_- = (1-  \Gamma_{\text{M5}})/2$ with
\bea
 \Gamma_{\text{M5}}&=& \frac1{5! \cL_\text{M5}}   \Gamma_0\Gamma^{N_1\cdots N_5} \varepsilon_{N_1\cdots N_5} |_{\text{M5}}~,
\eea
where $\cL_{\text{M5}} = \sqrt{g_\text{M5}}$ is the Dirac-Born-Infeld Lagrangian on the M5-brane,
and the subscript~$|_{\text{M5}}$ denotes the pullback to the worldvolume of the M5-brane.
This leads to a BPS bound
\bea
\| \cP_- \epsilon \|^2 = \epsilon^\dag \cP_- \epsilon \ge 0~,
\eea
which is saturated if and only if the probe M5-brane is supersymmetric. 
We rewrite this bound as
\bea\label{Bound5}
 \epsilon^\dag \epsilon \cL_{\text{M5}} \vol_5  &\geq&   \nu_5 |_\text{M5} ~,
\eea
where $\vol_5$
is the volume form on the spatial part of the worldvolume of the M5-brane,
and the five-form $\nu_5$ is defined as the bilinear
\bea
\nu_5 = \bar \epsilon \Gamma_{(5)}\epsilon~,
\eea
with $\bar \epsilon = \epsilon^\dag \Gamma_0$ and $\Gamma_{(n)} = \frac1{n!} \Gamma_{N_1\cdots N_n}\dd X^{N_1} \wedge \cdots\wedge \dd X^{N_n}$.

We will now use the analysis of general AdS$_4$ solutions in~\cite{Gabella:2012rc} to obtain the BPS condition for a probe M5-brane extending along all of AdS$_4$ and wrapping a 2d submanifold of $Y_7$.
The 11d metric is written as the warped product
$g_{11} = \lambda^{-1} (g_{\mathrm{AdS}_4}+g_{7})$ and the gamma matrices split accordingly as
\bea\label{Gammasplit}
\Gamma_\alpha &=& \rho_\alpha\otimes 1~, \qqq \Gamma_{a+3} \ = \ \rho_5\otimes \gamma_a~,
\eea
where $\alpha, \beta= 0,1,2,3$ and $a,b=1,\ldots,7$ are orthonormal frame indices for AdS$_4$ and $Y_7$ respectively: 
$\{\rho_\alpha,\rho_\beta\}=2\eta_{\alpha\beta}$, $\{\gamma_a,\gamma_b\}=2\delta_{ab}$.
We have defined the chirality matrix $\rho_5= \ii \rho_0\rho_1\rho_2\rho_3$. 
The 11d spinor $\epsilon$ splits into two spinors $\psi^+_1,\psi^+_2$ on AdS$_4$ and two internal spinors $\chi_1,\chi_2$ on $Y_7$:
\bea \label{11dspinorsplit}
\epsilon &= & \sum_{i=1,2} \psi_i^+ \otimes   \lambda^{-\frac14} \chi_i + (\psi_i^+)^c \otimes  \lambda^{-\frac14}   \chi_i^c~.
\eea
Given that $\epsilon^\dag \epsilon= 2 \lambda^{-\frac12} ( \| \psi_1^+\|^2 + \| \psi_2^+\|^2) $, the left-hand side of the BPS condition~\eqref{Bound5} becomes
\bea\label{LHSBPS}
\epsilon^\dag \epsilon \cL_{\text{M5}} \vol_5 = 2 \lambda^{-\frac72} ( \| \psi_1^+\|^2 + \| \psi_2^+\|^2)  \sqrt{g_\text{AdS$_4$}} \sqrt{g_2}\vol_5~,
\eea 
where $g_2$ stands for the determinant of the metric induced on the internal submanifold.
The right-hand side gives
\bea
\nu_5|_\text{M5}  &=&  2 \lambda^{-\frac72}  \sqrt{g_\text{AdS$_4$}}  \dd r  \wedge \dd\alpha_1 \wedge \dd\alpha_2 \wedge \sum_{i,j}  \Im \left[  (\psi_i^+)^\dag \psi_j^+ \  \bar \chi_i  \gamma_{(2)} \chi_j \right]_\text{M5} ~,
\eea
with $r ,\alpha_1,\alpha_2$ spatial coordinates on AdS$_4$ (see the AdS$_4$ metric in~\eqref{AdS4metric}).
We see that the 2d submanifold wrapped by the M5-brane is calibrated by the following two-form:
\bea\label{calibformVU}
\sum_{i,j}  \text{Im} \left[  (\psi_i^+)^\dag \psi_j^+ \  \bar \chi_i  \gamma_{(2)} \chi_j \right] &=&  \left( \| \psi_1^+\|^2 + \| \psi_2^+\|^2\right)  V_+ +   \left( \| \psi_1^+\|^2 - \| \psi_2^+\|^2\right) \Im U \nn
&&  + 2 \Re[ (\psi_1^+)^\dag \psi_2^+]  \Re U+ 2 \Im  [  (\psi_1^+)^\dag \psi_2^+ ] V_- ~,
\eea
where we used the notation
\bea
V_\pm &=& \frac{1}{2\ii} \left( \bar \chi_+\gamma_{(2)} \chi_+\pm \bar \chi_-\gamma_{(2)} \chi_-\right)~, \qqq
U= \bar \chi_+\gamma_{(2)} \chi_-~,
\eea
with $\chi_\pm =(\chi_1 \pm \ii \chi_2)/\sqrt{2} $.
Note that the AdS$_4$ spinors appear in the same combination in the left-hand side~\eqref{LHSBPS} of the BPS condition and in front of $V_+$ in the right-hand side~\eqref{calibformVU}.
This indicates that the M5-brane should be calibrated only by the term involving $V_+$, since the other terms in~\eqref{calibformVU} would generally lead to constraints on the AdS$_4$ spinors, thus breaking the $\cN=2$ supersymmetry that we wish to preserve (we confirm this argument by explicit calculations in appendix~\ref{appAdS4}).

In the case of an AdS$_4$ solution arising from wrapped M5-branes (no electric flux), the calibration form reads
\bea
V_+ = \lambda \sqrt{1-\lambda^3\rho^2} J_2~.
\eea
This means that the probe M5-brane does not wrap the circle $S^1_\text{R}$ parameterized by $\psi$ and so, in order to preserve the corresponding $U(1)$ R-symmetry, it must be located where this circle shrinks to a point, that is at
\bea
\boxed{\rho = 0}
\eea
We show in appendix~\ref{appAdS4} that this condition ensures that the pullbacks of the other two-forms appearing in~\eqref{calibformVU} vanish: $U |_\text{M5}  =V_- |_\text{M5} =  0$.

We remark that the condition that the probe M5-brane is calibrated essentially by $J_2$ at $\rho=0$ is consistent with its origin from a special Lagrangian in the UV.
Indeed, the $SU(3)$-structure of the Calabi-Yau 3-fold decomposes into the $SU(2)$-structure as
\bea
J = J_1 + \hat w\wedge \hat u~, \qqq \Omega = (J_3 + \ii J_2)\wedge (\hat w + \ii \hat u)~,
\eea
with $\hat u$ a unit one-form on CY$_3$, which reduces to $\lambda^{-1/2}\dd r$ at $\rho=0$ (see appendix C of~\cite{Gauntlett:2006ux}). 
It is then easy to see that, on the 3-cycle consisting of the internal 2d submanifold and the AdS$_4$ radial direction $r$, $J$ and $\Im \Omega$ restrict to zero, while $\Re\Omega$ gives a volume form.

Focusing on the PS solution reviewed in section~\ref{sec:AdS4}, we arrive at the following BPS condition for a supersymmetric probe M5-brane on AdS$_4$: 
\bea\label{BPSdetJ2}
\boxed{ \bar\vol_2   = \bar J_2\big|_\text{M5} }
\eea
Here $\bar\vol_2  = \sqrt{\bar g_2} \dd \tau \wedge \dd\sigma$ is the volume form on the internal part of the M5-brane worldvolume induced from the metric
\bea\label{metricT1H3}
\bar g = \frac{\dd x^2 + \dd y^2 + \dd z^2}{z^2} +\D Y^a \D Y^a~,
\eea
and the calibration two-form is given in terms of the vielbeins $\bar e = \{ \dd x, \dd y , \dd z\} /z$ by
\bea
\bar J_2 = \epsilon^{abc} Y^a \D Y^b \wedge \bar e^c~.
\eea

%%%%%%%%%%%%%%%%%%%%
\subsection{Conormal bundles}\label{conormal}
%%%%%%%%%%%%%%%%%%%%

We have derived that a supersymmetric probe M5-brane that extends along all of AdS$_4$ must be at $\rho=0$, where $S^1_\text{R}$ shrinks, and wrap a surface calibrated by the two-form $\bar J_2$ in the 5d space with metric $\bar g$ given in~\eqref{metricT1H3}.
In the next two subsections we will present some natural classes of solutions to the BPS condition~\eqref{BPSdetJ2}.
They all share the interesting feature that they appear to descend from conormal bundles in the cotangent bundle $T^*M_3$.

Recall that in the UV the M5-branes should be thought of as wrapping special Lagrangian submanifolds of $T^*M_3$.
The original stack of $N$ M5-branes is wrapping the 3-manifold $M_3$ itself, while there might be additional M5-branes that wrap other intersecting Lagrangians. A simple example of a Lagrangian submanifold of $T^*M_3$ is the conormal bundle $N^* L = \text{Ann}(TL)$ of a submanifold $L \subset M_3$. 
If $L$ is a knot in $M_3$, a well-known construction of the knot complement $M_3\backslash L$ consists in intersecting branes wrapped on $M_3$ and on $N^*L$ inside $T^*M_3$ (see~\cite{Ooguri:1999bv} and some recent applications in~\cite{Witten:2011zz,Aganagic:2013jpa}).

In the AdS$_4$ geometries that we study, what remains of the UV cotangent bundle $T^*M_3$ is the unit cotangent bundle $T^*_1M_3$ described by the metric $\bar g$ in~\eqref{metricT1H3} --- the radial direction in the cotangent fibers has been absorbed in the radial direction of AdS$_4$ and in $\rho$ (see section 5 in~\cite{Gauntlett:2006ux}).
Similarly, a Lagrangian conormal bundle $N^*L \subset T^*M_3$ descends to a Legendrian unit conormal bundle $N_1^*L\subset T^*_1M_3$.
The embedding of a probe M5-brane is therefore fully specified by the submanifold $L$ that it wraps in $M_3$.
The remaining position of the M5-brane on $S^2$ is then simply given by the fiber of $N_1^*L$.

Since the M5-brane is calibrated by the two-form
\bea
\bar J_2 = \epsilon^{abc} Y^a (\dd  Y^b + \omega^b{}_d  Y^d) \wedge \bar e^c~,
\eea
there are essentially two options for the submanifold $L\subset M_3$: it can be a curve or a surface.
We will analyze these two cases in turn in the next subsections.

For the moment, we forget about the quotient by the discrete subgroup $\Gamma\subset PSL(2,\CC)$ that produces the closed 3-manifold $M_3=\HH^3/\Gamma$, and study submanifolds $L \subset \HH^3$ of hyperbolic 3-space itself.
We will come back to the interpretation of our results in terms of the closed 3-manifold $M_3$ in section~\ref{hyperbolicinterpretation}.

Before proceeding we remark that the BPS condition~\eqref{BPSdetJ2} is invariant under any isometry $\gamma \in \text{Isom}^+(\HH^3) \cong PSL(2,\CC)$, combined with the corresponding $SO(3)$ transformation on the coordinates $Y^a$.
More explicitly, $\gamma$ induces a transformation $\cO \in SO(3)$ on the vielbeins, $\bar e^a \to \cO^a{}_b \bar e^b$, which extends to the constrained coordinates on $S^2$, $Y^a\to \cO^a{}_b Y^b$.
This invariance will allow us to focus first on solutions involving submanifolds $L \subset \HH^3$ that take particularly simple forms in the upper half-space model of hyperbolic 3-space, and then to obtain entire classes of solutions by acting with elements of $PSL(2,\CC)$.

%%%%%%%%%%%%%%%%%%%%%%%%%%%%
\subsection{Line defects in $\HH^3$}\label{codim2M5}
%%%%%%%%%%%%%%%%%%%%%%%%%%%%

\begin{figure}[t]
\centering
\includegraphics[width=\textwidth]{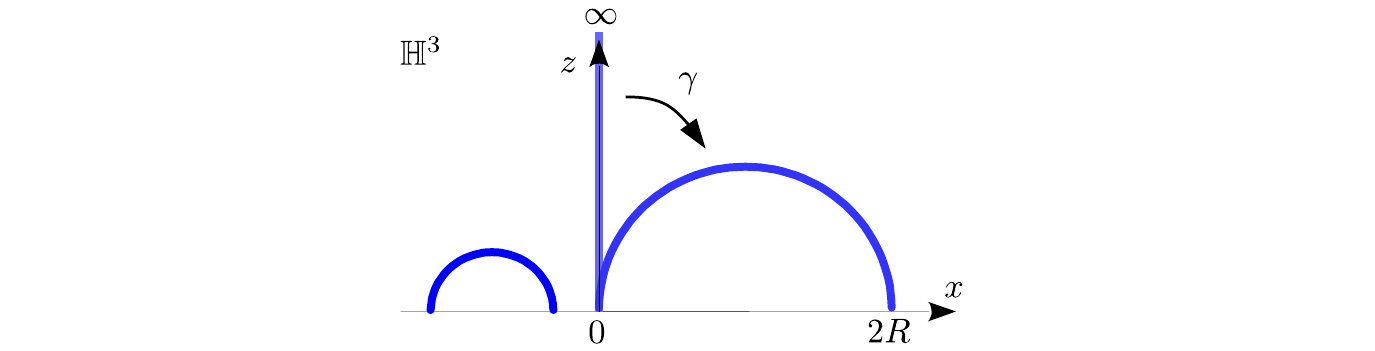}
\caption{Geodesics in the upper half-space model of $\HH^3$ are vertical lines and semicircles. They can be transformed into one another by elements $\gamma\in PSL(2,\CC)$.}
\label{geodesics12}
\end{figure}

\begin{figure}[tbh]
\centering
\includegraphics[width=\textwidth]{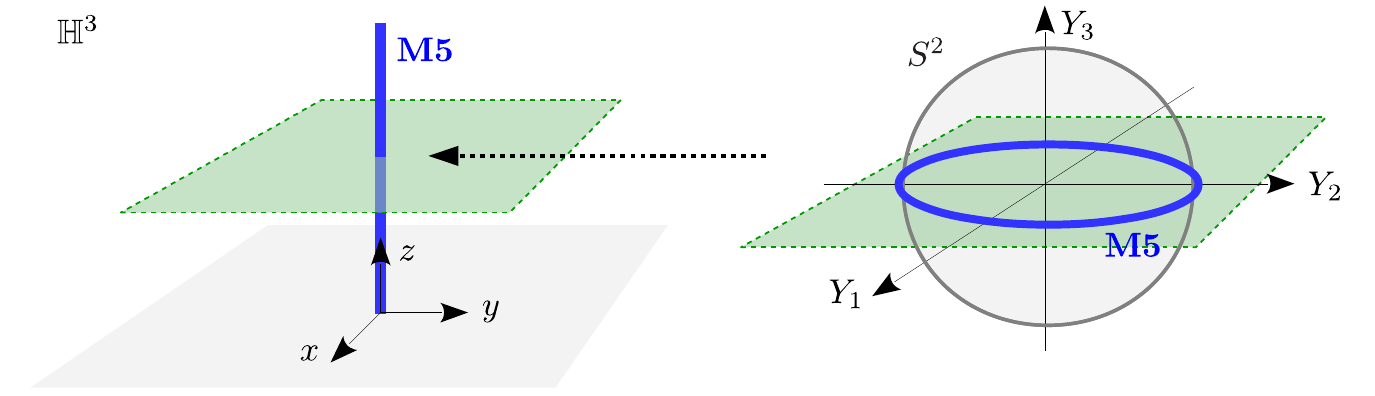}
\caption{\emph{Left}: BPS M5-brane on a straight geodesic in $\HH^3$, and a plane normal to it. \emph{Right}: The conormal space at any point of the straight geodesic intersects the $S^2$-fiber of the unit cotangent space $T^*_1\HH^3$ along the equator.}
\label{straightgeod}
\end{figure}

We now present a class of BPS probe M5-branes that extend along all of AdS$_4$ and wrap the unit conormal bundle $N_1^*L\subset T_1^*\HH^3$, with $L$ a geodesic curve in $\HH^3$. 
In the upper half-space model of hyperbolic 3-space, geodesics can be either vertical straight lines, or semicircles orthogonal to the boundary at $z=0$ (see figure~\ref{geodesics12}).
In the case where $L$ is a straight geodesic along the $z$-axis, the conormal bundle $N^*L$ intersects the $S^2$-fibers of the unit cotangent bundle $T^*_1\HH^3$ along the equator (see figure~\ref{straightgeod}).
Denoting the internal worldvolume coordinates by $\tau\in [0,\infty) $ and $\sigma\in [0,2\pi] $, we can write the embedding for this solution as
\bea
\{ x,y,z\} = \{0,0,\tau \}~, \qqq \{ Y_1,Y_2,Y_3\} = \{ \cos \sigma, \sin \sigma, 0\}~.
\eea

\begin{figure}[b]
\centering
\includegraphics[width=\textwidth]{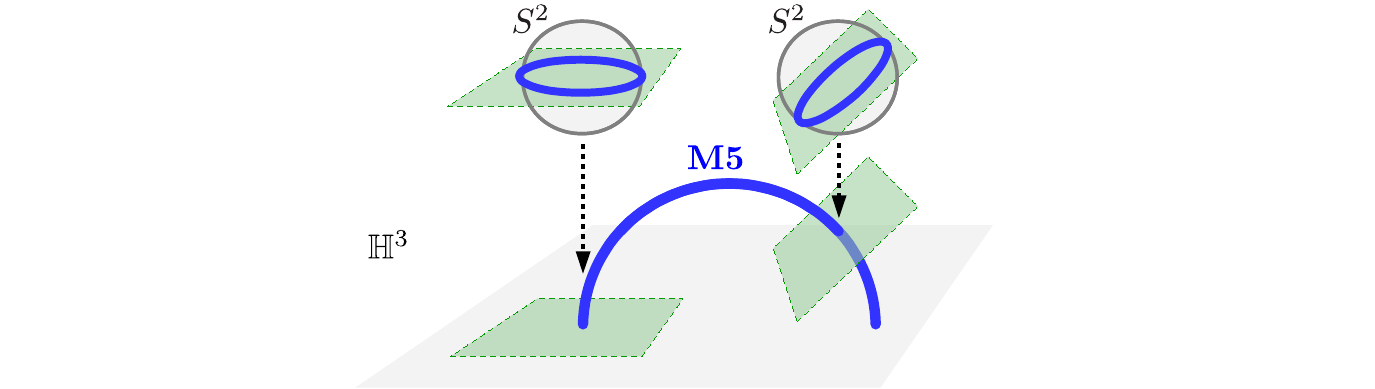}
\caption{BPS M5-brane on a geodesic semicircle in $\HH^3$. The conormal fiber intersects the $S^2$ along a great circle that rotates as the base point moves along the semicircle.}
\label{gaygeod}
\end{figure}

We can then generate solutions for any geodesic in $\HH^3$ by acting with transformations $\gamma \in PSL(2,\CC)$.
To obtain a semicircle geodesic we act on the straight geodesic with the transformation
\bea
\gamma = \begin{pmatrix} 0 & 1 \\ -1 & \frac1{2R} \end{pmatrix} ~. 
\eea
Applying the $PSL(2,\CC)$ action~\eqref{transfowz} we get
\bea
x' = \frac{2R}{1+4R^2 \tau^2}~, \qqq  y'=0 ~, \qqq  z' = \frac{4R^2 \tau}{1+4R^2 \tau^2}~,
\eea
which describes a semicircle centered at $\{x,y,z\}=\{R,0,0\}$ with radius $R$ (see figure~\ref{geodesics12}):
\bea 
(x' - R)^2 + (z')^2 =R^2~.
\eea
On the $S^2$-fibers, the M5-brane is now wrapping a great circle whose inclination depends on the base point (see figure~\ref{gaygeod}).
We can also shift the geodesic by acting with a parabolic element of $PSL(2,\CC)$, and rotate it with an elliptic element.

We will discuss in section~\ref{hyperbolicinterpretation} how such probe M5-branes on geodesics in $\HH^3$ can be interpreted as wrapping closed geodesics in $M = \HH^3 / \Gamma$. 
If we were to wrap an increasing number of M5-branes on a closed geodesic $L$, we would eventually produce a knot complement $M \backslash L$. 

%%%%%%%%%%%%%%%%%%%%%%%%%%%%
\subsection{Surface defects in $\HH^3$}

We now consider a supersymmetric probe M5-brane wrapping AdS$_4\times N_1^*L$, where $L$ is a surface in $\HH^3$.
We found two classes of solutions, namely surfaces that are equidistant from a point at infinity (horospheres) or from a geodesic in $\HH^3$ (tubes).

\subsubsection{Horospheres}\label{horosphere}

\begin{figure}[t]
\centering
\includegraphics[width=\textwidth]{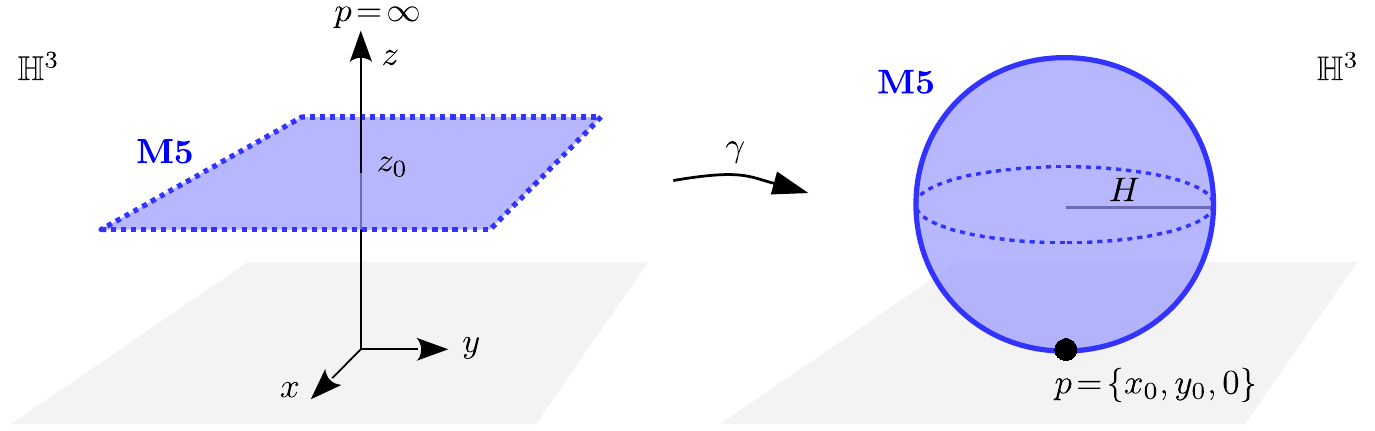}
\caption{\emph{Left:} BPS M5-brane on a horosphere around the point at infinity, which is simply a horizontal plane $z=z_0$. \emph{Right:} A transformation $\gamma \in PSL(2, \CC)$ can turn it into a horosphere around a point $p=\{x_0,y_0,0\}$, which is a Euclidean sphere tangent to $p$.}
\label{M5horo}
\end{figure}

Horospheres are surfaces that are equidistant from a point $p$ on the boundary $\del \HH^3 = S^2_\infty$ of hyperbolic 3-space. 
In the upper-half plane model of $\HH^3$, the point $p$ can be either on the plane $z=0$, in which case the horosphere is a Euclidean sphere tangent to $p$, or it can be at $\infty$, in which case the horosphere is a horizontal plane
(see figure~\ref{M5horo}).
The embedding corresponding to a horizontal plane is
\bea
\{ x,y,z\} = \{ \sigma_1, \sigma_2, z_0 \}~, \qqq \{Y_1, Y_2, Y_3 \} = \{ 0,0,\pm1\}~,
\eea
where $\{\sigma_1,\sigma_2\}\in \RR^2$ are worldvolume coordinates and $z_0$ is a constant.
On the $S^2$-fibers the M5-brane can be at the north and south poles.

Just like for the geodesic solution, we can then generate any horosphere by acting on this horizontal plane with some transformation $\gamma \in PSL(2,\CC)$.
We can parameterize the horosphere with center $p=\{x_0,y_0,0\}$ and radius $H$ as
\bea
\{ x,y,z\} &=& \{ x_0 + H\sin \sigma_1 \cos \sigma_2,y_0 + H\sin\sigma_1 \sin \sigma_2 , H +H\cos  \sigma_1\}~, 
\eea
where now $\sigma_1 \in [0,\pi]$ and $\sigma_2\in [0,2\pi]$.
The conormal fiber over a point of this horosphere intersects the $S^2$-fiber of $T^*_1 \HH^3$ at the corresponding point (or at the antipode): 
\bea
\{Y_1, Y_2, Y_3 \} = \pm \{ \sin\sigma_1  \cos  \sigma_2, \sin\sigma_1 \sin \sigma_2, \cos \sigma_1 \}~.
\eea

\subsubsection{Tubes}

\begin{figure}[t]
\centering
\includegraphics[width=\textwidth]{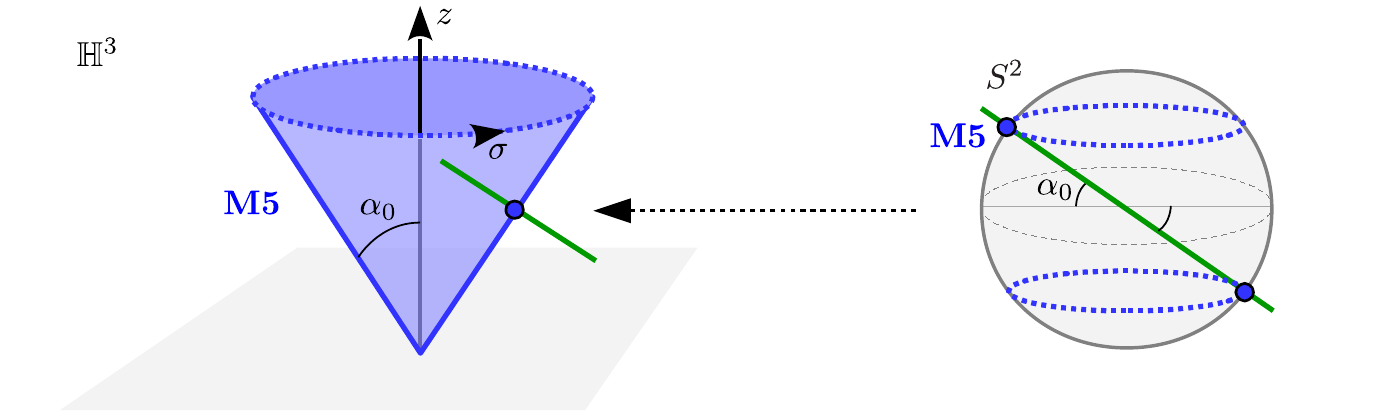}
\caption{\emph{Left}: BPS M5-brane on a cone with aperture $2\alpha_0$ around a vertical geodesic. \emph{Right}: The $S^2$-fiber over a point at $\sigma = \pi/2$  intersects the conormal fiber at the points $\{Y_1, Y_2, Y_3 \} =  \pm\{0, \cos \alpha_0 , -\sin  \alpha_0 \}$. For any point of the cone, the M5-brane is at a point on a horizontal circle (dashed) on the $S^2$-fiber.}
\label{M5cone}
\end{figure}

A BPS M5-brane can also wrap a surface that is equidistant from a geodesic in $\HH^3$, or in other words a tube.
In the case of a vertical geodesic, such an equidistant surface is simply given by a vertical cone with its apex on the plane $z=0$ (see figure~\ref{M5cone}). 
The cone with its apex at $\{x_0,y_0,0\}$ and with aperture $2\alpha_0 $ can be parameterized as
\bea
\{x,y,z\} = \{ x_0 + \tau \tan \alpha_0 \cos \sigma , y_0 + \tau \tan \alpha_0 \sin \sigma ,\tau\} ~,
\eea
with $\tau\in [0,\infty) $ and $\sigma\in [0,2\pi] $.
The conormal fiber over a point of the cone intersects $S^2$ at a point that lies on a horizontal circle of radius $\cos \alpha_0$:
\bea
\{Y_1, Y_2, Y_3 \} = \pm \{\cos \alpha_0  \cos  \sigma, \cos \alpha_0  \sin \sigma, -\sin  \alpha_0 \}~.
\eea

\begin{figure}[tbh]
\centering
\includegraphics[width=\textwidth]{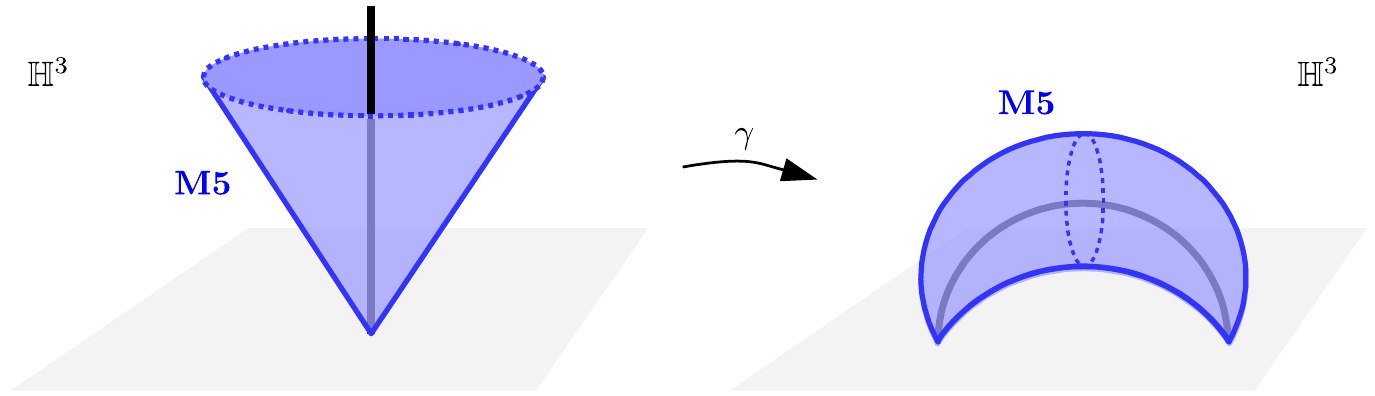}
\caption{A surface equidistant from a straight geodesic can be transformed by an element $\gamma \in PSL(2,\CC)$ into a surface equidistant from a geodesic semicircle, which looks like a banana.}
\label{M5banana}
\end{figure}

To obtain a surface equidistant from a semicircular geodesic, we again apply a transformation $\gamma \in PSL(2,\CC)$.
The resulting surface looks like a banana\footnote
{``Time flies like an arrow; fruit flies like a banana.'' (misattributed to Groucho Marx)} 
(see figure~\ref{M5banana}).

\begin{figure}[tbh]
\centering
\includegraphics[width=\textwidth]{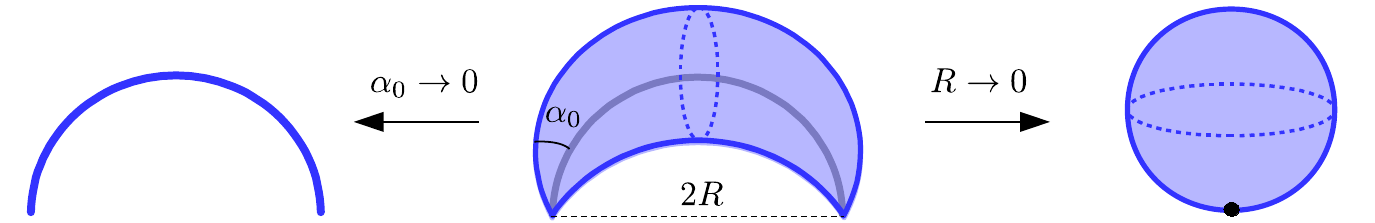}
\caption{The geodesic and horosphere solutions can be obtained as limits of the banana solution.}
\label{M5bananalimits}
\end{figure}

Note that this banana solution contains the other solutions that we found as special limits.
Indeed, if we send the aperture $\alpha_0$ of a banana to zero, we obtain a geodesic as in section~\ref{codim2M5}, while if we bring the two endpoints of a banana together, we obtain a horosphere as in section~\ref{horosphere} (see figure~\ref{M5bananalimits}).
Thus the space of BPS probe M5-branes is connected.
%

%%%%%%%%%%%%%%%%%%%
\subsection{Extra $U(1)$ symmetry} 

For all the solutions that we found (geodesic, horosphere, tube) the metric induced on the 2d submanifold wrapped by the M5-brane turns out to be flat. 
Interestingly, a theorem of Sasaki says that any complete flat surface in $\HH^3$ is either a horosphere or a surface equidistant from a geodesic~\cite{sasaki}. These are precisely the two classes of solutions that we found for an M5-brane on a surface in $\HH^3$. 

An important consequence of the fact that their embeddings are of the form AdS$_4\times \RR \times S^1$ is that the probe M5-branes preserve an extra $U(1)$ symmetry --- in addition of course to the $U(1)$ R-symmetry associated with $S^1_{\text R}$.

In fact, all our solutions can be alternatively derived without thinking about conormal bundles, but by imposing an ansatz that preserves a $U(1)$ symmetry acting simultaneously on $\HH^3$ and $S^2$.
Denoting the worldvolume coordinates by $\tau \in \RR_+$ and $\sigma\in [0,2\pi]$, we write the ansatz as
\bea
\{x,y,z\} &=& \{ f(\tau) \cos \sigma, f(\tau) \sin \sigma, a_0 \tau + z_0 \}~, \nn
\{Y_1, Y_2, Y_3 \} &=& \pm \{\cos \alpha_0  \cos  \sigma, \cos \alpha_0  \sin \sigma, -\sin  \alpha_0 \}~.
\eea
where $a_0,z_0,\alpha_0$ are constants and $f(\tau)$ is a function. 
Inserting this into the BPS condition~\eq{BPSdetJ2} gives a differential equation for $f(\tau)$: 
\bea
&& 2 \cos ^2\alpha_0 \left[-(\cos 2 \alpha_0-3) f ^2 +2 \sin 2 \alpha_0 z f  +(\cos 2 \alpha_0+3)  z^2\right] \left(f'\right)^2 \nn
&&  + 2 a_0 \cos \alpha_0 \left[ (\sin 3 \alpha_0-7 \sin \alpha_0) f ^2+2 (\cos 3 \alpha_0-3 \cos \alpha_0) z f -4 \sin \alpha_0 \cos ^2\alpha_0  z^2\right] f' \nn
&&  +a_0^2 \left[- \left(\sin ^22 \alpha_0+2 \cos 2 \alpha_0-6\right) f^2+ 8 \sin ^3\alpha_0 \cos \alpha_0 z f   + \sin ^22 \alpha_0  z^2 \right]  =0~.
\eea
We find three types of solutions, which reproduce the simple ones that we presented above:
\bea
&&\text{Straight geodesic:}  \qqq f(\tau)=0~,\quad a_0=1~, \quad z_0=0~, \quad \alpha_0 =0~, \nn
&&\text{Horizontal plane:}  \qqq f(\tau)=\tau~,\quad a_0=0~, \quad z_0 > 0~, \quad \alpha_0 = \pi/2 ~, \\
&&\text{Vertical cone:}  \qqq   f(\tau)= \tau \sin \alpha_0 ~,\quad a_0= \cos \alpha_0 ~, \quad z_0=0~, \quad \alpha_0 \in [0,\pi/2]~. \nonumber
\eea
Taking the limit $\alpha_0\to 0 $ of the vertical cone gives the straight geodesic, while for $\alpha_0\to \pi/2$ the cone coincides with the horizontal plane with $z_0\to 0$. 
We can then produce general geodesics, horospheres, or bananas by acting with elements of $PSL(2,\CC)$ in $\HH^3$ and with the corresponding $SO(3)$ transformations on the coordinates $Y^a$.

%%%%%%%%%%%%%%%%%%%%%%%%%%%%
\section{M2-branes ending on M5-branes}\label{secM2}
%%%%%%%%%%%%%%%%%%%%%%%%%%%%

Probe M2-branes wrapping 2d internal submanifolds correspond to BPS operators in the dual 3d $\cN=2$ gauge theory.
In the presence of probe M5-branes, we can consider a probe M2-brane on submanifolds with a 1d boundary on probe M5-branes.
The BPS condition implies that M2-branes are located at $\rho=0$ and are calibrated by the two-form $J_3$.
We will describe M2-branes wrapping the $S^2$-fiber over a point in $\HH^3$, as well as M2-branes wrapping a hemisphere in $\HH^3$. 
Such embeddings can end on M5-branes wrapping geodesics, but not on M5-branes wrapping surfaces in $\HH^3$.

%%%%%%%%%%%%%%%%%%%%%%%%%%%%
\subsection{BPS condition}\label{secM2BPS}

Similar arguments to the ones reviewed in section~\ref{secBPSM5} lead to a BPS bound for a supersymmetric probe M2-brane:
\bea\label{Bound2}
\epsilon^\dag \epsilon \cL_\text{M2} \vol_2  \geq    \bar\epsilon \Gamma_{(2)} \epsilon|_\text{M2}~.
\eea
Using the decomposed gamma matrices~\eqref{Gammasplit} and spinor~\eqref{11dspinorsplit}, we find 
\bea
 \bar\epsilon \Gamma_{(2)} \epsilon &=& 4 \lambda^{-2}\Re \left[  \bar \psi_1^+ (\psi_2^+)^c \bar \chi_1 \gamma_{(2)} \chi_2^c \right] ~.
\eea
This leads to the BPS condition
\bea\label{BPScondM2}
\left(\|\psi_1^+\|^2 +\|\psi_2^+\|^2 \right) \sqrt{g_2} \vol_2 &=&  2  \Re[\bar \psi_1^+ (\psi_2^+)^c ] \sqrt{1-\lambda^3 \rho^2}  \lambda J_3  \nn 
&&    - 2 \Im [\bar \psi_1^+ (\psi_2^+)^c  ] \left( \lambda J_1 - \frac{\lambda^{7/2}\rho}{4\sqrt{1- \lambda^3 \rho}} \dd\rho \wedge \hat w \right)  ~.
\eea
If the probe M2-brane is calibrated by the first term, the condition on the AdS$_4$ spinors is
\bea
\|\psi_1^+\|^2 +\|\psi_2^+\|^2  = 2  \Re[\bar \psi_1^+ (\psi_2^+)^c] ~,
\eea
which imposes that the M2-brane is at the center of AdS$_4$, that is at $r =0$, as can be seen from the explicit expressions of the spinor bilinears given in~\eqref{psibilinears} and~\eqref{bilinearM2}.
On the other hand, if the probe M2-brane is calibrated by the second term in~\eqref{BPScondM2}, the AdS$_4$ condition cannot be solved for finite $r $, and we therefore exclude this case.

Since the M2-brane is essentially calibrated by the two-form $J_3$ given in~\eqref{ansatzstr},
it does not wrap  $S^1_\text{R}$, and so, in order to preserve the corresponding $U(1)$ R-symmetry, it must be located at $\rho=0$, where $S^1_\text{R}$ shrinks.
This is also the location of the probe M5-branes on which we want the M2-brane to end.

The BPS condition for a supersymmetric probe M2-brane in the PS solution is then
\bea\label{BPSdetJ3}
\boxed{\bar \vol_2 = \bar J_3\big|_\text{M2} }
\eea
where $\bar \vol_2 = \sqrt{\bar g_2} \dd\sigma_1 \wedge \dd\sigma_2$ is the volume form on the internal part of the M2-brane worldvolume induced from the metric~\eqref{metricT1H3}, and the calibration two-form is
\bea
\bar J_3 =  \tfrac12 \epsilon^{abc} Y^a \left(   \D Y^b \wedge \D Y^c - \bar e^b\wedge \bar e^c \right)~.
\eea
We see that there are in principle three types of embeddings to consider: the M2-brane could wrap a point, a line, or a surface in $\HH^3$.
There are no solutions to the BPS condition for an M2-brane on a line, but we found BPS M2-branes at points and on surfaces in $\HH^3$.

%%%%%%%%%%%%%%%
\subsection{Spherical M2-branes}
\begin{figure}[tbh]
\centering
\includegraphics[width=\textwidth]{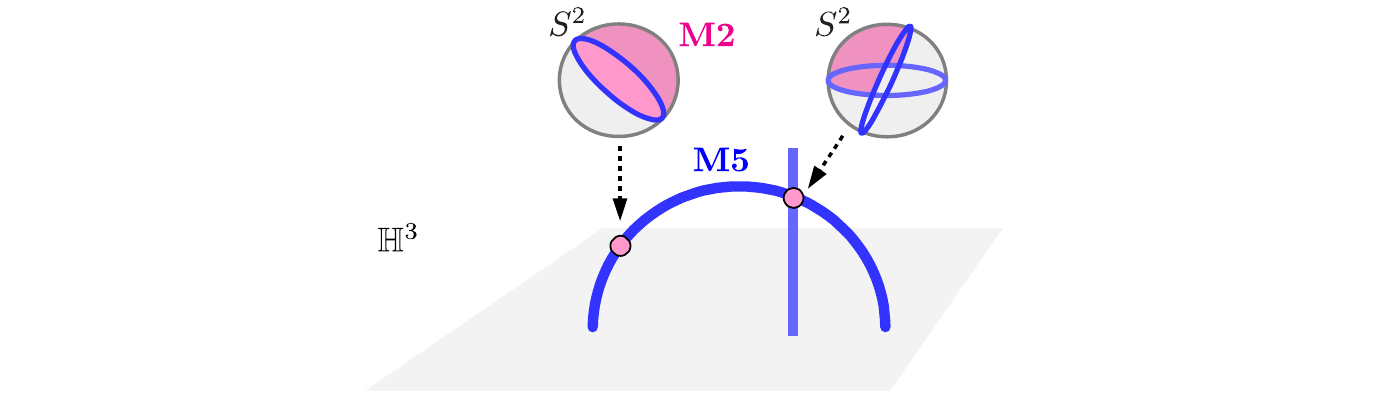}
\caption{\emph{Left}: BPS M2-brane at a point of a geodesic in $\HH^3$ wrapped by an M5-brane. In the $S^2$-fiber over this point, the M2-brane can end on the great circle wrapped by the M5-brane. \emph{Right}:~The M2-brane can sit at an intersection between geodesic M5-branes and stretch between their great circles in the $S^2$-fiber.}
\label{M2point}
\end{figure}

If the M2-brane sits at a constant point in $\HH^3$, the BPS condition does not impose any constraints on the coordinates $Y^a$ on the $S^2$-fiber in $T^*_1\HH^3$:
\bea
\{x,y,z\} = \{ x_0,y_0, z_0 \}~, \qqq \{Y_1,Y_2,Y_3\}  \quad \text{not constrained.}
\eea
Without any additional probe M5-brane, the M2-brane would simply wrap the whole $S^2$.
However, there is a new possibility in the presence of an M5-brane that wraps a geodesic in $\HH^3$ and a great circle on the $S^2$: the M2-brane can sit at a constant point on the geodesic, and wrap an hemisphere of the $S^2$-fiber that ends on the great circle.
We can also consider a probe M2-brane sitting at the intersection of two geodesics wrapped by M5-branes.
On the $S^2$-fiber over the intersection point, the M2-brane stretches between the two great circles wrapped by the M5-branes (see figure~\ref{M2point}).

In contrast, an M2-brane at a point in $\HH^3$ cannot end along a 1d boundary on an M5-brane wrapping a surface in $\HH^3$ (horosphere or tube) because such an M5-brane is point-like on the $S^2$-fiber.

%%%%%%%%%%%%%%%
\subsection{Hyperbolic M2-branes}

A BPS M2-brane can wrap a geodesic plane in $\HH^3$, which is either a vertical plane, or a hemisphere ending on the boundary $z=0$. Note that geodesic planes are copies of $\HH^2$ inside $\HH^3$.

\begin{figure}[t]
\centering
\includegraphics[width=\textwidth]{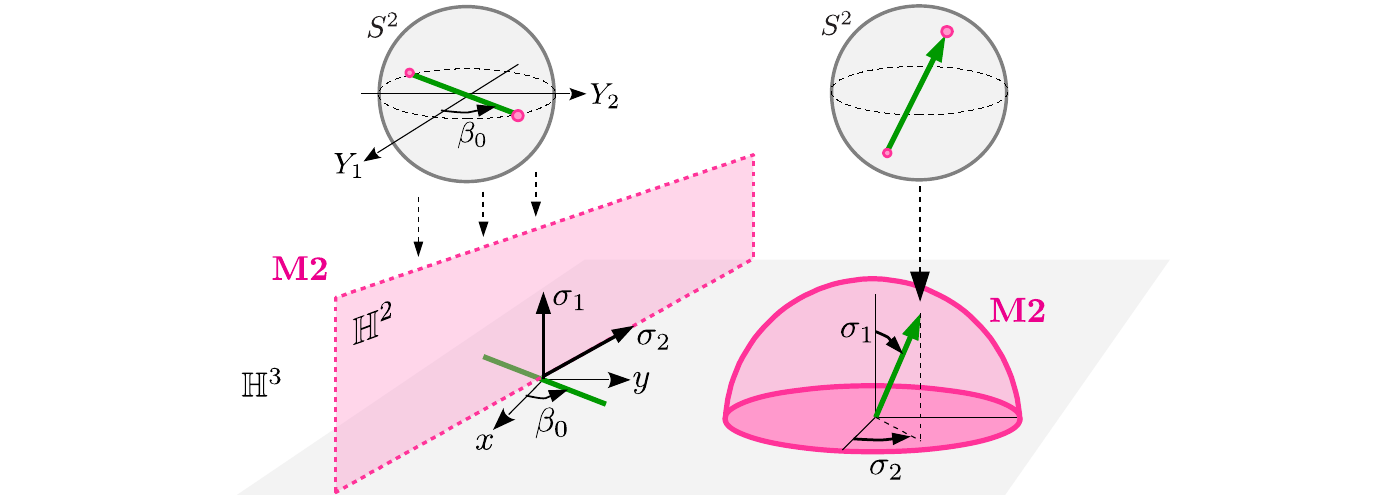}
\caption{\emph{Left}: BPS M2-brane on a vertical plane in $\HH^3$, and at a constant point on $S^2$. \emph{Right}:~BPS M2-brane on a hemisphere in $\HH^3$, and at the corresponding point on the $S^2$-fiber.}
\label{M2hemi}
\end{figure}

In the case of a vertical plane, the M2-brane is at a constant point on the equator of $S^2$ depending on the orientation $\beta_0$ of the plane: 
\bea
\{x,y,z\} &=& \{ x_0 -\sigma_2 \sin \beta_0 ,y_0+ \sigma_2 \cos \beta_0, \sigma_1 \}~, \nn 
\{Y_1,Y_2,Y_3\} &=& \pm \{\cos \beta_0,\sin \beta_0,0\}~,
\eea
with $\sigma_1 \in \RR_+$ and $\sigma_2 \in \RR$.
For a hemisphere, in each $S^2$-fiber over a point of the hemisphere the M2-brane is located at the corresponding point (or at the antipode):
\bea
\{x,y,z\} &=& \{ x_0 +R_0 \sin \sigma_1 \cos \sigma_2 ,y_0+R_0 \sin \sigma_1 \sin \sigma_2, R_0 \cos \sigma_1 \}~, \nn 
\{Y_1,Y_2,Y_3\} &=&  \pm\{\sin \sigma_1 \cos \sigma_2,\sin \sigma_1 \sin \sigma_2,\cos \sigma_1\}~,
\eea
with now $\sigma_1 \in [0, \pi/ 2]$ and $\sigma_2 \in [0, 2\pi]$.
These two configurations are shown in figure~\ref{M2hemi}.

\begin{figure}[tbh]
\centering
\includegraphics[width=\textwidth]{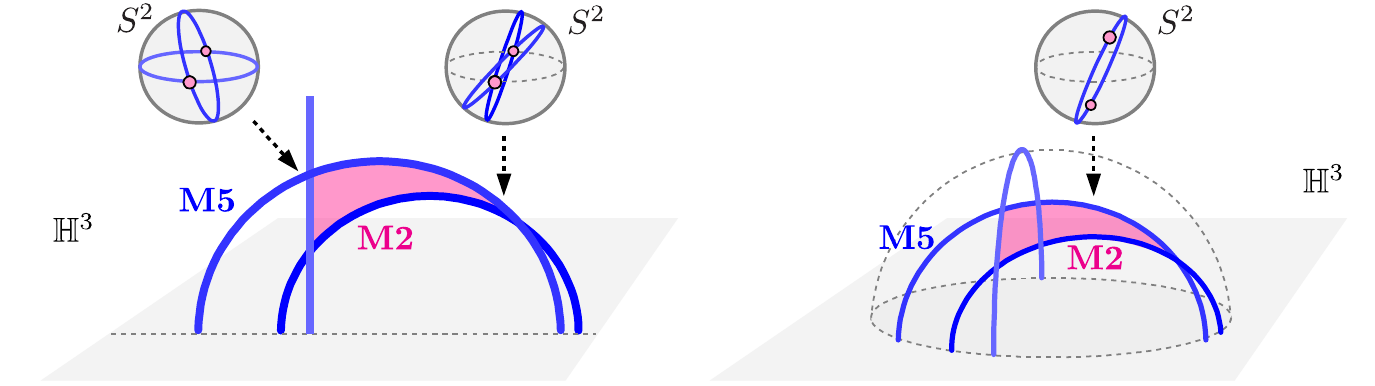}
\caption{\emph{Left}: BPS M2-brane stretched between M5-branes on geodesics in the same geodesic plane. On the $S^2$-fibers, the M2-brane is at a constant point, around which the great circles wrapped by the M5-branes rotate. \emph{Right}:~A similar configuration on a hemisphere.}
\label{M2stretch}
\end{figure}

It is amusing to note that, just like for the M5-branes, the submanifolds wrapped by the M2-branes are unit conormal bundles in $T_1^*\HH^3$.
If a geodesic curve is embedded inside a geodesic plane (vertical plane or hemisphere), the conormal fiber of the curve automatically contains the conormal fiber of the plane.
We can thus consider M2-branes stretching between geodesic M5-branes in the same geodesic plane (see figure~\ref{M2stretch}).

Although it would naively appear that a hemispheric M2-brane can have a 1d boundary on a horosphere or on a tube wrapped by an M5-brane, closer inspection reveals that the M2-brane and the M5-brane are located at different points on the $S^2$-fibers, and hence do not genuinely meet.

%%%%%%%%%%%%%%%%%%%%%%%%%%%%
\section{From geodesics to knot complements}\label{hyperbolicinterpretation}
%%%%%%%%%%%%%%%%%%%%%%%%%%%%

In this section we propose a potential interpretation of our results in terms of the geometry of hyperbolic 3-manifolds (see appendix~\ref{hyperbolicgeom} for a small review).

We started with the Pernici-Sezgin AdS$_4$ solution, which involves a \emph{closed} hyperbolic 3-manifold $M_3 = \HH^3/\Gamma$.
However, in the analysis of sections~\ref{secM5} and~\ref{secM2} we neglected the quotient by $\Gamma$ and obtained BPS embeddings for probe M5-branes and M2-branes in $T^*_1\HH^3$.
We found in particular that a BPS M5-brane can wrap a geodesic in $\HH^3$, that is a vertical line or semicircle with its endpoints on $S^2_\infty$.
We would like to understand how to think about this geodesic after the quotient to $M_3$.

\begin{figure}[tbh]
\centering
\includegraphics[width=\textwidth]{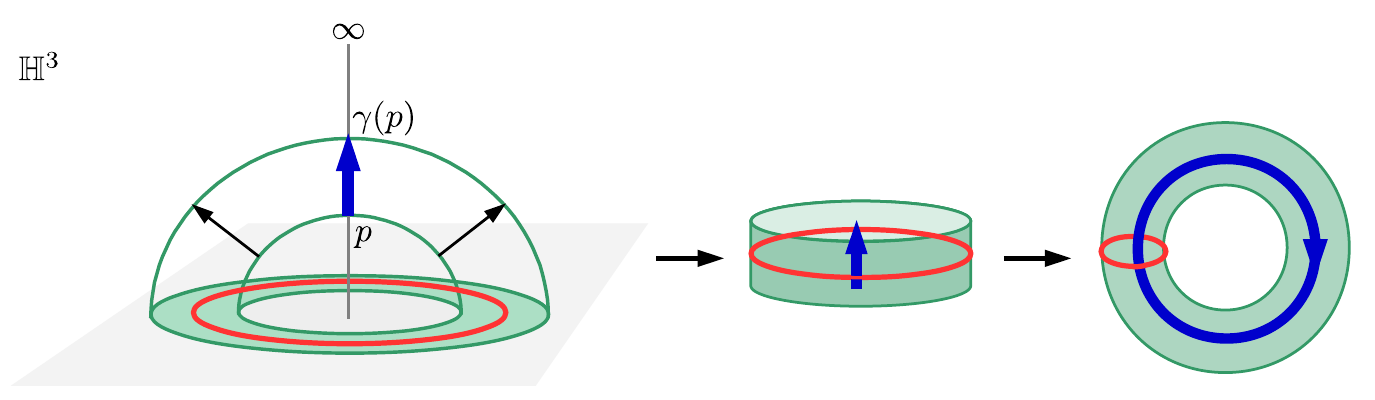}
\caption{Solid torus $\HH^3/\langle \gamma \rangle$, where $\gamma$ is a loxodromic transformation (here with real eigenvalue). The corresponding closed geodesic is shown in blue. A meridian on the boundary of the solid torus is shown in red.}
\label{solidtorus}
\end{figure}

In the closed 3-manifold $M_3=\HH^3/\Gamma$, we expect that probe M5-branes wrap closed geodesics, and these correspond to \emph{loxodromic} elements $\gamma \in \Gamma$.
To see this, recall that a loxodromic transformation acts as a screw motion (rotation plus translation) around its axis, which is the unique geodesic in $\HH^3$ between its two fixed points on $S^2_\infty$. 
A point $p$ on this axis is simply translated along the axis by $\gamma$, and so the segment $[p, \gamma(p)]$ projects to a closed geodesic in $\HH^3/\langle \gamma \rangle$ (see figure~\ref{solidtorus}).
It is thus natural to associate geodesic M5-branes to loxodromic elements in $\Gamma$.

\begin{figure}[t]
\centering
\includegraphics[width=\textwidth]{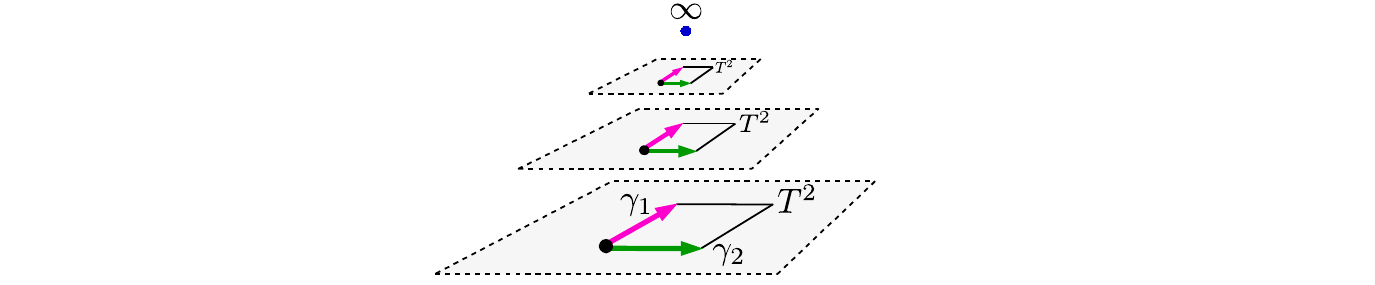}
\caption{A pair of parabolic elements $\gamma_1,\gamma_2 \in \Gamma$ with common fixed point (here at $\infty$) turn the surrounding horospheres (horizontal planes) into nested cusp tori.}
\label{toruscusp}
\end{figure}

One of our motivations for studying the addition of a probe M5-brane on a closed geodesic in $M_3$ is that this can be seen as the first step towards the creation of the complement $M_3\backslash \cK$ of a knot $\cK\subset M_3$.
The knot $\cK$ that has been removed from $M_3$ should be thought of as being located at a fixed point on $S^2_\infty$ that is shared by two parabolic elements $\gamma_1,\gamma_2\in \Gamma$.
This parabolic fixed point, or \emph{cusp}, is in a sense ``stretched'' into the knot.
The parabolic group $\langle \gamma_1, \gamma_2\rangle$ turns the horospheres around the fixed point into nested tori, so that the cusp neighborhood looks like $T^2 \times [0,\infty)$ (see figure~\ref{toruscusp}).\footnote
{This should be compared with the cusp neighborhood $S^1\times [0,\infty)$ of a puncture on a Riemann surface.}
A knot complement $M_3\backslash \cK$ is thus a non-compact manifold with a cusp torus surrounding the knot at infinity.
To produce such a drastic deformation of $M_3$ we would certainly need to wrap a very large number of coincident M5-branes on a closed geodesic.

It turns out that there is a well-known procedure in hyperbolic geometry, called \emph{conformal Dehn surgery}, which precisely generates a cusp torus from a sequence of loxodromic elements \cite{Jorgensen} (see also the video {\it Not Knot}~\cite{notknot}).\footnote{We thank Roland van der Veen for pointing out this wonderful video.}
It consists in drilling out a tubular neighborhood of a closed geodesic in $M_3$, and then filling it in with increasingly twisted solid tori.
In the limit of infinite twisting, the solid torus ``fractures'' and creates a cusp torus on $S^2_\infty$. 
It is then natural to interpret the addition of coincident M5-branes on a closed geodesic as the operation of conformal Dehn surgery, with the twist number related to the number of M5-branes.

%%%%%%%%%%%
\subsection{Conformal Dehn surgery}
We illustrate conformal Dehn surgery with an explicit example~\cite{Jorgensen} (see section 4.9 in~\cite{marden} and also chapter 9 in~\cite{thurston}).
Consider first a non-compact cusped hyperbolic 3-manifold $\HH^3/\Gamma$, where $\Gamma$ is a parabolic group of rank 2:
\bea
\Gamma = \langle \gamma_1(w) = w + 1, \gamma_2(w) = w + \tau \rangle~,
\eea
with $w=x+\ii y$ and $\Im \tau >0$. 
There is a cusp torus $T^2= \CC/\Gamma$ at the boundary at infinity.
We choose a pair of simple loops $\{\alpha, \beta\}$ on $T^2$ that intersect once, so that every simple loop can be expressed in the form $p\alpha + q \beta$, with $p$ and $q$ relatively prime integers.
Performing $(p,q)$-Dehn surgery means gluing a solid torus $S^1\times D^2$ to $T^2$ such that the curve $p\alpha + q \beta$ matches the meridian on boundary of the solid torus.
In particular, $(1,0)$-Dehn surgery applied to a knot complement $M_3 \backslash \cK$ just gives the closed manifold $M_3$.

We will now perform $(1,q)$-Dehn surgery and see how the cusp torus is recreated in the limit $q\to \infty$.
The suitable solid torus (see figure~\ref{solidtorus}) is 
\bea
\left( \HH^3 \cup   \CC \right)  / \langle L_q \rangle~,
\eea
where $L_q\subset PSL(2,\CC)$ is the loxodromic transformation
\bea
L_q (u) =  \lambda_q u   ~, \qqq \text{with} \qquad \lambda_q = \exp\left( - \frac{2 \pi \ii   \tau}{1+ q \tau}\right) ~.
\eea
The boundary of the solid torus, $T^2_q =  \CC / \langle L_q \rangle$, is identified with $T^2$ via the conformal map
\bea
u_q : \quad T^2 & \to & T_q^2~, \nn
w &\mapsto  &  u_q(w) = \exp \left(- \frac{2\pi\ii w}{1 + q\tau} \right)~.
\eea
Indeed, we see that the generators $\gamma_1$ and $\gamma_2$ producing $T^2$ can be expressed in terms of the generator $L_q$ on $T^2_q$:
\bea\label{propuq1}
(u_q \circ \gamma_1 )(w) =  (L_q{}^{-q} \circ u_q) (w)~, \qqq
(u_q \circ \gamma_2 )(w) =  (L_q \circ u_q) (w)~.
\eea
Note that if we change the basis on $T^2$ via $\gamma_1(w) \to \gamma_{1,q}(w) =  w+(1+q \tau)$, which corresponds to replacing $\alpha$ by $\alpha + q \beta$, we get
\bea\label{propuq2}
(u_q \circ \gamma_{1,q} )(w) &=&  u_q (w)~.
\eea
This means that the image of $\alpha + q \beta$ is a meridian on the solid torus.
A straight line with tangent vector $1+q \tau$ is mapped by $u_q$ to a circle centered at 0, which then projects to a meridian on $T^2_q $ (see figure~\ref{solidtorus}).

\begin{figure}[t]
\centering
\includegraphics[width=\textwidth]{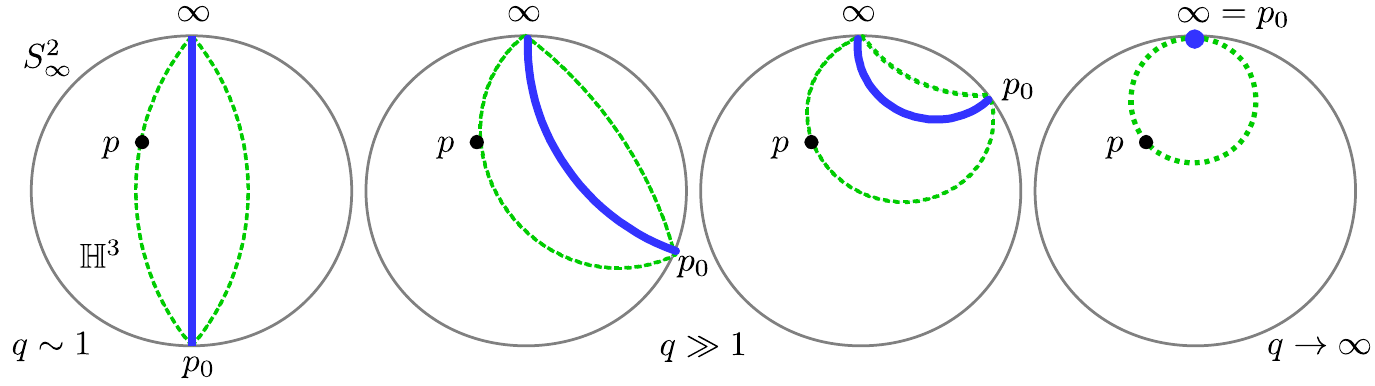}
\caption{Sequence of axes (blue) for the loxodromic elements $\{\tilde L_q\}$ in the Poincar\' e ball model of $\HH^3$. The point $p\subset\HH^3$ is translated by $\tilde L_q$ on an invariant surface (dashed) surrounding the axis. For finite $q$, this surface looks like a banana. As $q\to \infty$, the fixed point $p_0$ merges with the other fixed point at $\infty$, and the invariant surface becomes a horosphere.}
\label{bananasequence}
\end{figure}

\begin{figure}[h]
\centering
\includegraphics[width=\textwidth]{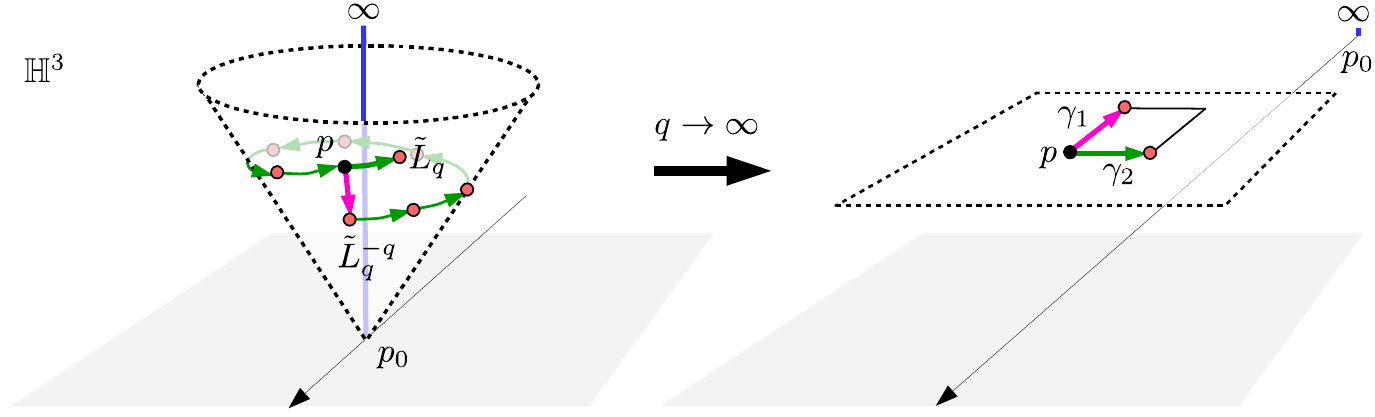}
\caption{\emph{Left:} Acting with the loxodromic transformations $\tilde L_q$ or $\tilde L_q^{-q}$ on a point $p$ translates it in different directions on a cone around the axis of $\tilde L_q$. \emph{Right:} in the limit $q\to \infty$, the fixed point $p_0$ recedes to $\infty$ as in figure~\ref{bananasequence}, and the cone becomes a horizontal plane (the horosphere around~$\infty$). We are left with two parabolic elements $\gamma_1=\lim \tilde L_q^{-q}$ and $\gamma_2= \lim\tilde L_q$.}
\label{loxocone}
\end{figure}

However, given that $\lambda_q \to 1$ for $q\to \infty$, the sequence of loxodromic transformations $\{L_q\}$ converges to the identity.
In order to obtain a sequence that converges geometrically to $\Gamma= \langle \gamma_1 , \gamma_2 \rangle$, we conjugate $L_q$ such that the fixed points are at $p_0=\tau/(1-\lambda_q)$ and $\infty$:
\bea
\tilde L_q (u)= \tilde \gamma_q L_q \tilde \gamma_q^{-1}(u) = \lambda_q u + \tau~, \qqq \text{with}\qquad \tilde \gamma_q(u) = u + p_0~.
\eea
We now have $\lim_{q\to \infty} \tilde L_q (u) =  u + \tau$.
Note also that the fixed point $p_0$ tends to the other fixed point at $\infty$ as $q\to \infty$, and the axis of $\tilde L_q$ contracts to a point (see figure~\ref{bananasequence}).
The conformal map is taken to be 
\bea
\tilde u_q(w) = p_0 \left(1 - u_q(w)\right)~,
\eea
which also satisfies properties such as~\eqref{propuq1} and~\eqref{propuq2}.
Finally, from the fact that $\lim\tilde u_q(w) = w$ it follows that
\bea
\lim_{q\to \infty}  \tilde L_q{}^{-q}    = \gamma_1 ~, \qqq \lim_{q\to \infty}  \tilde L_q  =   \gamma_2 ~.
\eea
The sequence of loxodromic groups is said to converge \emph{geometrically} to the rank-2 parabolic group $\Gamma= \langle \gamma_1 , \gamma_2 \rangle$ (see figure~\ref{loxocone}).
In conclusion, the sequence of solid tori $\HH^3 / \langle \tilde L_q \rangle$ converges for $q\to \infty$ to the non-compact 3-manifold $\HH^3/ \Gamma$ with a cusp at its core. 
%

%%%%%%%%%%%%%%%%%%%%
\section{Discussion}\label{SECdiscussion}
%%%%%%%%%%%%%%%%%%%%

We have presented supersymmetric probe M5-branes in the Pernici-Sezgin  AdS$_4$ solution that preserve the superconformal symmetries of the dual $\cN=2$ SCFT.
They extend along all of AdS$_4$ and are located where the $S^1_\text{R}$ dual to the R-symmetry shrinks to a point.
We have shown that the BPS condition for M5-branes (but also for M2-branes ending on them) then boils down to the calibration of a surface in the unit cotangent bundle $T^*_1M_3$ (with $S^2$-fibers) of the hyperbolic 3-manifold $M_3 = \HH^3/\Gamma$.
We have found solutions corresponding to all the natural objects appearing in the geometry of hyperbolic 3-manifolds.
BPS M5-branes can wrap geodesic curves in $M_3$, on which BPS M2-branes wrapping geodesic surfaces can end. 
M5-branes can also wrap invariant surfaces which are equidistant from the axis of a loxodromic or elliptic transformation (tubes), or from a parabolic fixed point (cusp torus or annulus).

In all these cases, the calibrated surface turns out to be simply the unit conormal bundle $N^*_1L\subset T^*_1M_3$ of a submanifold $L\subset M_3$.
For example, an M5-brane on a geodesic in $M_3$ also wraps a rotating great circle in the $S^2$-fibers.
This suggests that in the UV regime the M5-branes are wrapping special Lagrangian submanifolds given by the conormal bundles $N^*L\subset T^*M_3$.
It would be very interesting to make this perspective more precise and to study the flow from the UV to the IR.

If we were to wrap a large number of supersymmetric M5-branes on $N_1^*L$, they would ultimately backreact on the Pernici-Sezgin geometry and produce a new AdS$_4$ solution, arising from intersecting stacks of M5-branes (this would be an AdS$_4$ analogue of the general AdS$_5$ solution found in~\cite{Lin:2004nb}).
Just like in conformal Dehn surgery, we expect that the original closed 3-manifold $M_3$ will develop cusps along the way, and could then be a knot (or link) complement.
Since the BPS M5-branes that we found preserve an extra $S^1$ in addition to $S^1_\text{R}$, we predict that this general AdS$_4$ solution will have a $U(1)^2$ isometry.

An important open problem is to match our results to the dual 3d $\cN=2$ SCFTs.
Gauge theories associated with 3-manifolds have been constructed in various ways, starting with~\cite{Dimofte:2011ju,Cecotti:2011iy}, and for higher rank in~\cite{Dimofte:2013iv}.
Theories associated with closed 3-manifolds have been presented in~\cite{Gadde:2013sca,Chung:2014qpa}.
The volume of the hyperbolic 3-manifold $M_3$ will give the free energy of the 3d theory on an ellipsoid~\cite{Martelli:2011fu} (see~\cite{Gang:2014qla} for calculations for knot complements).

The probe M5-branes on geodesics are expected to correspond to flavor symmetries. In particular, SCFTs associated with knot complements should have an $SU(N)$ flavor symmetry~\cite{Dimofte:2013iv}.
Probe M2-branes ending on M5-branes correspond to BPS operators,
and the volumes of the surfaces they wrap will give their conformal dimensions.
We can anticipate that M5-branes on geodesics and M5-branes on surfaces will play very different roles in the 3d theory, since M2-branes can end on the former but not on the latter.
We have found two types of M2-branes, which will correspond to two types of BPS operators: those stretching between geodesics on a hyperbolic surface, or those stretching between great circles on an $S^2$-fiber.
On the other hand, M5-branes on tubes in $M_3$ might find an interpretation in terms of domain walls and couplings to 4d $\cN=2$ theories, as for example in~\cite{Dimofte:2013lba}.

We have seen that the worldvolume of a BPS M5-brane is of the form AdS$_4\times \RR\times S^1$.
The Kaluza-Klein reduction on this $S^1$ of the two-form potential on the worldvolume of a probe M5-brane produces a $U(1)$ gauge field in AdS$_4$, corresponding to a global $U(1)$ symmetry in the dual 3d theory.
This extra $U(1)$ symmetry, which comes in addition to the $U(1)$ corresponding to the R-symmetry, was recently shown in~\cite{Chung:2014qpa} to play a key role in the 3d-3d correspondence.

\

%%%%%%%%%%
\acknowledgments

We would like to thank Mina Aganagic, Clay Cordova, Tudor Dimofte, Oliver Fabert, Sergei Gukov, Daniel Jafferis, Juan Maldacena, Christoph Schweigert, and Nick Warner for useful discussions.
We are grateful to the Simons Center for Geometry and Physics for hospitality during part of this project. IB is supported in part by the DOE grant DE-FG03-84ER-40168, ANR grant 08-JCJC-0001-0. The work of MG is supported by the German Science Foundation (DFG) within the Research Training Group 1670 ``Mathematics Inspired by String Theory and QFT.''

%%%%
\newpage
%%%%%%%%%%%%%%%%%%%%%%%%%%%%%%
%%%%%%%%%%%%%%%%%%%%%%%%%%%%%%
\appendix
%%%%%%%%%%%%%%%%%%%%%%%%%%%%%%
%%%%%%%%%%%%%%%%%%%%%%%%%%%%%%

%%%%%%%%%%%%%%%
\section{Hyperbolic 3-manifolds}\label{hyperbolicgeom}
%%%%%%%%%%%%%%%

We review some relevant aspects of the geometry of hyperbolic 3-manifolds (good references are for example~\cite{marden,matsuzaki,benedetti,ratcliffe}).

\begin{figure}[h]
\centering
\includegraphics[width=\textwidth]{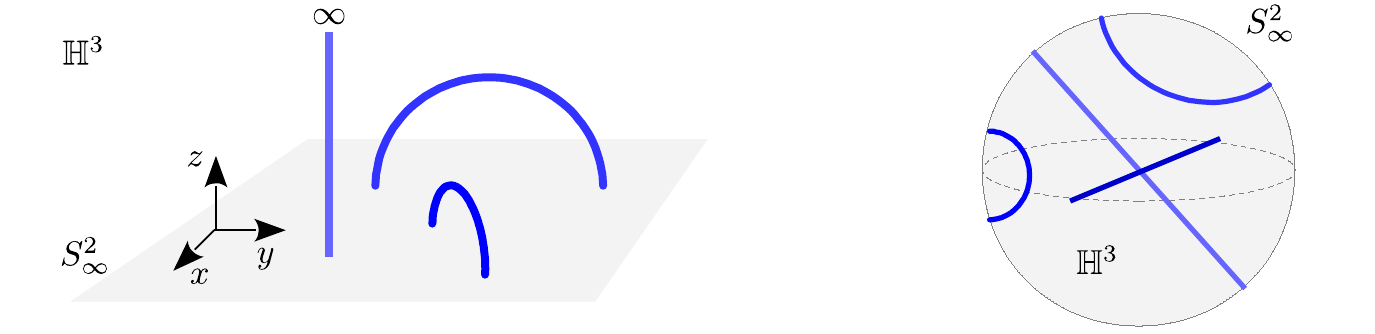}
\caption{Geodesics in the upper half-space model of $\HH^3$ (\emph{left}) and in the ball model (\emph{right}).}
\label{H3geod}
\end{figure}

There are several commonly used models of hyperbolic 3-space $\HH^3$.
We mostly use the upper half-space model with the metric
\bea
\dd s^2 (\HH^3) &=& \frac{\dd x^2 + \dd y^2 + \dd z^2 }{z^2}~, \qqq z>0~.
\eea
We also refer occasionally to the ball model $\{ \vec x \in \RR^3 : |\vec x| <1  \}$ with the metric
\bea
\dd s^2(\HH^3) = \frac{4 |\dd\vec x|^2 }{(1-|\vec x|^2)^2}~.
\eea
We denote the boundary at infinity of $\HH^3$ by $S^2_\infty$, which in the case of the upper-half space model has to be understood as the plane $z=0$ together with the point at infinity, $\CC \cup \{\infty\}$.
In the upper half-space model, geodesics are vertical lines and semicircles orthogonal to $S^2_\infty$, while they are diameters and arcs orthogonal to $S^2_\infty$ in the ball model (see figure~\ref{H3geod}).
Geodesic surfaces are vertical half-planes and hemispheres orthogonal to $S^2_\infty$.

\begin{figure}[h]
\centering
\includegraphics[width=\textwidth]{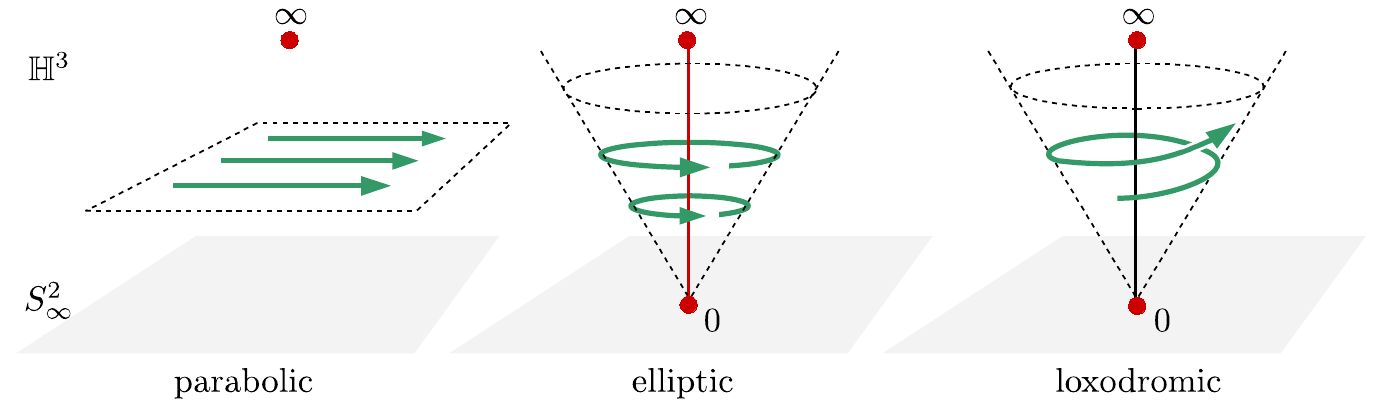}
\caption{The three types of $PSL(2,\CC)$ transformations. A \emph{parabolic} transformation acts as a translation, an \emph{elliptic} one as a rotation around its axis, and a \emph{loxodromic} one as a screw motion along its axis. The dashed surfaces (horizontal planes or cones) are left invariant.}
\label{TransfoTypes}
\end{figure}

The group of orientation-preserving isometries of $\HH^3$ is $\text{Isom}^+(\HH^3) \cong PSL(2,\CC)$.
The action of $PSL(2,\CC)$ on $\HH^3$ can be expressed as fractional linear transformation on a quaternion $q=x+\ii y + \text{j} z$: 
\bea
q \mapsto \gamma(q) = \frac{a q + b}{c q + d}~, \qqq \text{with}\qquad \gamma = \begin{pmatrix}  a&b \\ c&d \end{pmatrix}  \in PSL(2,\CC)~.
\eea
In terms of the complex coordinate $w= x+\ii y$ this gives
\bea\label{transfowz}
w \mapsto \frac{(aw+b)(\bar c \bar w + \bar d) + a \bar c z^2 }{|c w + d|^2 + |c|^2z^2}~, \qqq
z \mapsto \frac{z}{|c w + d|^2 + |c|^2z^2}~.
\eea
Any element of $PSL(2, \CC)$ is conjugate to one of the following three standard matrices:
\begin{itemize}
\item 
$\begin{pmatrix}  1&1 \\ 0&1 \end{pmatrix}$:
\qqq $\{w,z\} \mapsto \{w+1,z\}$ \qqq\qqq\qquad \;\, (\emph{parabolic}),
\item
$\begin{pmatrix}  \ex^{\ii\theta} & 0 \\ 0&\ex^{-\ii\theta} \end{pmatrix}$: 
\qquad $\{w,z\} \mapsto \{\ex^{2 \ii\theta}w,z\}$~, \; with $\theta < 2\pi$  \qqq\; (\emph{elliptic}),
\item
$\begin{pmatrix}  \lambda &0 \\ 0&\lambda^{-1} \end{pmatrix}$:
\qquad $\{w,z\} \mapsto \{\lambda^2 w,|\lambda|^2 z\}$~,  \; with $|\lambda|>1$  \qqq (\emph{loxodromic}).
\end{itemize}
A \emph{parabolic} transformation acts as a translation and has one fixed point on $S^2_\infty$ (the standard parabolic matrix fixes the point at infinity).
An \emph{elliptic} transformation acts as a rotation around the geodesic that connects its two fixed points on $S^2_\infty$ (0 and $\infty$ for the standard matrix); note that an elliptic transformation also fixes a curve inside $\HH^3$, that is its axis of rotation.
A \emph{loxodromic} transformation acts as a screw motion, rotating around its axis as well as translating along it, from the repelling to the attracting fixed point on $S^2_\infty$; the axis is mapped to itself (invariant) but not fixed.
These three types of transformations are illustrated in figure~\ref{TransfoTypes}.

An elliptic or loxodromic transformation leaves invariant a family of surfaces equidistant from its axis.
When the axis is a vertical line, such a surface is a cone centered on it; if the axis is a semicircle, the surface looks like a banana.
A parabolic transformation leaves invariant the family of surfaces called \emph{horospheres}, which are equidistant from its fixed point on $S^2_\infty$. 
When the fixed point is at $\infty$, horospheres are horizontal Euclidean planes, otherwise they are Euclidean spheres tangent to $S^2_\infty$ at the fixed point.

\

A hyperbolic 3-manifold $M_3$ can be represented as a quotient of hyperbolic space $\HH^3$ by a discrete subgroup $\Gamma \subset PSL(2,\CC)$, called the \emph{Kleinian group}:
\bea
M_3 = \HH^3 / \Gamma~.
\eea
Note that $\Gamma$ is a holonomy representation of the fundamental group $\pi_1(M_3)$ into $PSL(2,\CC)$.
If $\Gamma$ is torsion-free (no elliptic element), $M_3$ is an oriented manifold (possibly with boundary) with a complete hyperbolic structure.
On the other hand, if $\Gamma$ contains elliptic elements, $M_3$ is called a hyperbolic 3-orbifold, and the hyperbolic structure has conical singularities along the projection of the fixed rotation axes to $M_3$.

Parabolic elements in $\Gamma$ generate \emph{cusps}, which make $M_3$ non-compact.
If a fixed point is shared by a pair of parabolic elements, its cusp neighborhood is isometric to $T^2 \times [0,\infty)$, where $T^2$ is a torus generated by the pair of translations on a horosphere around the fixed point (as in figure~\ref{toruscusp}).
A parabolic fixed point that is not shared is associated with a cusp annulus that relates two punctures on the boundary of $M_3$.

%%%%%%%%%%%%%%
\section{AdS$_4$ Killing spinors}\label{appAdS4}

In this appendix we construct a pair of chiral Killing spinors $\psi^+_1$ and $\psi^+_2$ in AdS$_4$.
We write the AdS$_4$ metric in global coordinates as
\bea\label{AdS4metric}
\dd s^2 (\text{AdS}_4) = \frac14 \left( - \cosh^2 r  \dd t^2 + \dd r ^2 + \sinh^2 r  \dd\Omega_2^2 \right)~,
\eea
with the round metric on $S^2$ given by
$\dd\Omega_2^2 = \dd \alpha_1^2 + \sin^2 \alpha_1 \dd\alpha_2$. 
A (non-chiral) Killing spinor $\psi$ satisfies the equation
$\D_\mu \psi =   \rho_\mu \psi$,
or more explicitly
\bea
\left( \del_\mu + \frac14 \omega_\mu{}^{\alpha\beta} \rho_{\alpha\beta} \right) \psi = e^\alpha_\mu \rho_\alpha \psi~,
\eea
where $\omega^{\alpha\beta}$ is the spin connection, $e^\alpha$ are the vielbeins, and the matrices $\rho_\alpha$ satisfy the Clifford algebra $\{\rho_\alpha, \rho_\beta\} = 2 \eta_{\alpha\beta}$.
The general solution to this equation can be written as
\bea
\psi = \exp\left(\frac r  2 \rho_1\right) \exp\left(\frac t 2 \rho_0\right)\exp\left(\frac{\alpha_1} 2 \rho_{12}\right)\exp\left(\frac{\alpha_2} 2 \rho_{23}\right)\psi_0~,
\eea
with $\psi_0$ a constant spinor.
We can now project on the positive-chirality part $\psi^+ = \frac12 ( 1+ \rho_5) \psi$, so that $\rho_5 \psi^+ = \psi^+$, where the chirality matrix is $\rho_5 = \ii \rho_0\rho_1\rho_2\rho_3$. 
This chiral Killing spinor satisfies
$\D_\mu \psi^+ = \rho_\mu ( \psi^+)^c$,
with the superscript $c$ denoting charge conjugation (which inverts the chirality).

To perform explicit calculations, we choose a basis for the Clifford matrices:
\bea
\rho_0 = \ii \begin{pmatrix} 0 & \mathbb{I}_2 \\ \mathbb{I}_2 & 0 \end{pmatrix}~, \qqq \rho_a =\ii \begin{pmatrix} 0 & \sigma_a \\  - \sigma_a & 0 \end{pmatrix}~, \qqq \rho_5 =  \begin{pmatrix} -\mathbb{I}_2 & 0 \\  0  & \mathbb{I}_2 \end{pmatrix}~,
\eea
with the Pauli matrices $\sigma_a$ for $a=1,2,3$.
Charge conjugation then acts as $\psi^c = -\ii\rho_2\psi^*$.
A convenient choice for the two chiral Killing spinors used in the main text is
\bea
\psi^+_i =    \frac12 ( 1+ \rho_5)  \exp\left(\frac r  2 \rho_1\right) \exp\left(\frac t 2 \rho_0\right)\exp\left(\frac{\alpha_1} 2 \rho_{12}\right)\exp\left(\frac{\alpha_2} 2 \rho_{23}\right)\psi_{0i}~,
\eea
with
$\psi_{01} = (0,\ii,1,0)^\text{T}$ and $\psi_{02} = (-\ii,0,0,1)^\text{T}$.

The combinations of AdS$_4$ spinors that appeared in the calibration two-form~\eqref{calibformVU} for a supersymmetric probe M5-brane are then expressed as
\bea\label{psibilinears}
\| \psi_1^+\|^2 + \| \psi_2^+\|^2 &=&  2 \cosh r  ~, \nn
\| \psi_1^+\|^2 - \| \psi_2^+\|^2 &=& 2\sinh r  (\cos t\cos \alpha_1 + \sin t \sin \alpha_1 \sin \alpha_2)  ~, \nn
  (\psi_1^+)^\dag \psi_2^+  &=&  \sinh r  (\sin t\cos \alpha_1 - \cos t \sin \alpha_1 \sin \alpha_2)   ~.
\eea
Expanding the BPS condition~\eqref{Bound5} in powers of $r $, we see that the M5-brane must be calibrated by the term involving $V_+$ as claimed in section~\ref{secBPSM5}, and that the pullback of $U$ to its worldvolume must vanish.
We saw that the first condition requires the M5-brane to be at $\rho=0$ in order to preserve the $U(1)$ R-symmetry, which then implies the second condition
$U |_\text{M5}  = 0$ since we have 
\bea
U &=&  S^* \Re[\bar \chi_+^c \gamma_{(2)} \chi_-] + L^* \wedge P \nn
&=& \lambda^2  \ex^{-\ii(\psi-\tau)} \left[ - \lambda^{1/2}\rho J_1 + \frac14 \left( \frac1{ \sqrt{1-\lambda^3 \rho^2}} \dd \rho  - \ii \rho\sqrt{1-\lambda^3 \rho^2}  \dd\psi\right) \wedge \hat w \right]~.
\eea
Note that the other two-form appearing in~\eqref{calibformVU} is given by
\bea
V_- &=& \frac{\lambda^3 \rho}{16} \dd \rho \wedge \dd\psi~,
\eea
which also vanishes at $\rho=0$.

We also need the following spinor bilinear that appears in the BPS condition~\eqref{BPScondM2} for a probe M2-brane:
\bea\label{bilinearM2}
\bar \psi^+_1 (\psi_2^+)^c =  1+ \ii \sinh r  \sin \alpha_1 \cos \alpha_2~.
\eea

%%%%%%%%%%%%%%%%%%%%%%%%%%%%%%%%%%%%%%%
%%%%%%%%%%%%%%%%%%%%%%%%%%%%%%%%%%%%%%%
%\providecommand{\href}[2]{#2}
\begingroup\raggedright
\endgroup


\begin{thebibliography}{10}

\bibitem{Dimofte:2011ju}
T.~Dimofte, D.~Gaiotto, and S.~Gukov, {\it {Gauge Theories Labelled by
  Three-Manifolds}},  {\em Commun.Math.Phys.} {\bf 325} (2014) 367--419,
  [\href{http://xxx.lanl.gov/abs/1108.4389}{{\tt arXiv:1108.4389}}].

\bibitem{Cecotti:2011iy}
S.~Cecotti, C.~Cordova, and C.~Vafa, {\it {Braids, Walls, and Mirrors}},
  \href{http://xxx.lanl.gov/abs/1110.2115}{{\tt arXiv:1110.2115}}.

\bibitem{Dimofte:2013iv}
T.~Dimofte, M.~Gabella, and A.~B. Goncharov, {\it {K-Decompositions and 3d
  Gauge Theories}},  \href{http://xxx.lanl.gov/abs/1301.0192}{{\tt
  arXiv:1301.0192}}.

\bibitem{Lee:2013ida}
S.~Lee and M.~Yamazaki, {\it {3d Chern-Simons Theory from M5-branes}},  {\em
  JHEP} {\bf 1312} (2013) 035, [\href{http://xxx.lanl.gov/abs/1305.2429}{{\tt
  arXiv:1305.2429}}].

\bibitem{Cordova:2013cea}
C.~Cordova and D.~L. Jafferis, {\it {Complex Chern-Simons from M5-branes on the
  Squashed Three-Sphere}},  \href{http://xxx.lanl.gov/abs/1305.2891}{{\tt
  arXiv:1305.2891}}.

\bibitem{Pernici:1984nw}
M.~Pernici and E.~Sezgin, {\it {Spontaneous Compactification of
  Seven-dimensional Supergravity Theories}},  {\em Class.Quant.Grav.} {\bf 2}
  (1985) 673.

\bibitem{Gaiotto:2009gz}
D.~Gaiotto and J.~Maldacena, {\it {The Gravity duals of N=2 superconformal
  field theories}},  {\em JHEP} {\bf 1210} (2012) 189,
  [\href{http://xxx.lanl.gov/abs/0904.4466}{{\tt arXiv:0904.4466}}].

\bibitem{Bah:2013wda}
I.~Bah, M.~Gabella, and N.~Halmagyi, {\it {Punctures from Probe M5-Branes and
  N=1 Superconformal Field Theories}},
  \href{http://xxx.lanl.gov/abs/1312.6687}{{\tt arXiv:1312.6687}}.

\bibitem{Ooguri:1999bv}
H.~Ooguri and C.~Vafa, {\it {Knot invariants and topological strings}},  {\em
  Nucl.Phys.} {\bf B577} (2000) 419--438,
  [\href{http://xxx.lanl.gov/abs/hep-th/9912123}{{\tt hep-th/9912123}}].

\bibitem{Witten:2011zz}
E.~Witten, {\it {Fivebranes and Knots}},
  \href{http://xxx.lanl.gov/abs/1101.3216}{{\tt arXiv:1101.3216}}.

\bibitem{Aganagic:2013jpa}
M.~Aganagic, T.~Ekholm, L.~Ng, and C.~Vafa, {\it {Topological Strings, D-Model,
  and Knot Contact Homology}},  \href{http://xxx.lanl.gov/abs/1304.5778}{{\tt
  arXiv:1304.5778}}.

\bibitem{Acharya:2000mu}
B.~S. Acharya, J.~P. Gauntlett, and N.~Kim, {\it {Five-branes wrapped on
  associative three cycles}},  {\em Phys.Rev.} {\bf D63} (2001) 106003,
  [\href{http://xxx.lanl.gov/abs/hep-th/0011190}{{\tt hep-th/0011190}}].

\bibitem{Gauntlett:2000ng}
J.~P. Gauntlett, N.~Kim, and D.~Waldram, {\it {M Five-branes wrapped on
  supersymmetric cycles}},  {\em Phys.Rev.} {\bf D63} (2001) 126001,
  [\href{http://xxx.lanl.gov/abs/hep-th/0012195}{{\tt hep-th/0012195}}].

\bibitem{Gauntlett:2006ux}
J.~P. Gauntlett, O.~A. Mac~Conamhna, T.~Mateos, and D.~Waldram, {\it {AdS
  spacetimes from wrapped M5 branes}},  {\em JHEP} {\bf 0611} (2006) 053,
  [\href{http://xxx.lanl.gov/abs/hep-th/0605146}{{\tt hep-th/0605146}}].

\bibitem{Gabella:2012rc}
M.~Gabella, D.~Martelli, A.~Passias, and J.~Sparks, {\it {${\cal N}=2$
  supersymmetric AdS$_{4}$ solutions of M-theory}},  {\em Commun.Math.Phys.}
  {\bf 325} (2014) 487--525, [\href{http://xxx.lanl.gov/abs/1207.3082}{{\tt
  arXiv:1207.3082}}].

\bibitem{Martelli:2011fu}
D.~Martelli, A.~Passias, and J.~Sparks, {\it {The gravity dual of
  supersymmetric gauge theories on a squashed three-sphere}},  {\em Nucl.Phys.}
  {\bf B864} (2012) 840--868, [\href{http://xxx.lanl.gov/abs/1110.6400}{{\tt
  arXiv:1110.6400}}].

\bibitem{Gang:2014qla}
D.~Gang, N.~Kim, and S.~Lee, {\it {Holography of Wrapped M5-branes and
  Chern-Simons theory}},  {\em Phys.Lett.} {\bf B733} (2014) 316--319,
  [\href{http://xxx.lanl.gov/abs/1401.3595}{{\tt arXiv:1401.3595}}].

\bibitem{Becker:1995kb}
K.~Becker, M.~Becker, and A.~Strominger, {\it {Five-branes, membranes and
  nonperturbative string theory}},  {\em Nucl.Phys.} {\bf B456} (1995)
  130--152, [\href{http://xxx.lanl.gov/abs/hep-th/9507158}{{\tt
  hep-th/9507158}}].

\bibitem{Martelli:2003ki}
D.~Martelli and J.~Sparks, {\it {G structures, fluxes and calibrations in M
  theory}},  {\em Phys.Rev.} {\bf D68} (2003) 085014,
  [\href{http://xxx.lanl.gov/abs/hep-th/0306225}{{\tt hep-th/0306225}}].

\bibitem{sasaki}
S.~Sasaki, {\it {On complete flat surfaces in hyperbolic 3-space}},  {\em Kodai
  Mathematical Seminar Reports} {\bf 25} (1973) 449--457.

\bibitem{Jorgensen}
T.~Jorgensen and A.~Marden, {\it {Algebraic and Geometric Convergence of
  Kleinian Groups}},  {\em Math.~Scand.} {\bf 66} (1990) 47--72.

\bibitem{notknot}
C.~Gunn and D.~Maxwell, {\em Not Knot}.
\newblock AK Peters, 1991.
\newblock Video available on YouTube.

\bibitem{marden}
A.~Marden, {\em Outer Circles: An Introduction to Hyperbolic 3-Manifolds}.
\newblock Cambridge University Press, 2007.

\bibitem{thurston}
W.~P. Thurston, {\em The Geometry and Topology of Three-Manifolds}.
\newblock http://library.msri.org/books/gt3m/, 1980.

\bibitem{Lin:2004nb}
H.~Lin, O.~Lunin, and J.~M. Maldacena, {\it {Bubbling AdS space and 1/2 BPS
  geometries}},  {\em JHEP} {\bf 0410} (2004) 025,
  [\href{http://xxx.lanl.gov/abs/hep-th/0409174}{{\tt hep-th/0409174}}].

\bibitem{Gadde:2013sca}
A.~Gadde, S.~Gukov, and P.~Putrov, {\it {Fivebranes and 4-manifolds}},
  \href{http://xxx.lanl.gov/abs/1306.4320}{{\tt arXiv:1306.4320}}.

\bibitem{Chung:2014qpa}
H.-J. Chung, T.~Dimofte, S.~Gukov, and P.~Sulkowski, {\it {3d-3d Correspondence
  Revisited}},  \href{http://xxx.lanl.gov/abs/1405.3663}{{\tt
  arXiv:1405.3663}}.

\bibitem{Dimofte:2013lba}
T.~Dimofte, D.~Gaiotto, and R.~van~der Veen, {\it {RG Domain Walls and Hybrid
  Triangulations}},  \href{http://xxx.lanl.gov/abs/1304.6721}{{\tt
  arXiv:1304.6721}}.

\bibitem{matsuzaki}
K.~Matsuzaki and M.~Taniguchi, {\em Hyperbolic Manifolds and Kleinian Groups}.
\newblock Oxford University Press, 1998.

\bibitem{benedetti}
R.~Benedetti and C.~Petronio, {\em Lectures on Hyperbolic Geometry}.
\newblock Springer, 1992.

\bibitem{ratcliffe}
J.~G. Ratcliffe, {\em Foundations of Hyperbolic Manifolds}.
\newblock Springer, 2006.

\end{thebibliography}
\end{document}